\documentclass[a4paper,fleqn,usenatbib]{mnras}

\usepackage{times}
\usepackage[T1]{fontenc}
\usepackage{ae,aecompl}

\usepackage{graphicx}
\usepackage{amsmath}
\usepackage{amssymb}
\usepackage{aas_macros}
\usepackage{bm}

\defcitealias{pf10}{PF10}

\title[Grid vs. SPH on the small-scale turbulent dynamo]{A comparison between grid and particle methods on the small-scale dynamo in magnetised supersonic turbulence}

\author[Tricco, Price \& Federrath]{
Terrence S. Tricco,$^{1,2,3}$\thanks{E-mail: ttricco@cita.utoronto.ca}
Daniel J. Price$^{3}$ and
Christoph Federrath$^{4,3}$ \\
$^{1}$Canadian Institute for Theoretical Astrophysics, University of Toronto, 60 St. George Street, Toronto, ON  M5S 3H8, Canada\\
$^{2}$School of Physics, University of Exeter, Stocker Road, Exeter EX4 4QL, United Kingdom\\
$^{3}$Monash Centre for Astrophysics, School of Physics \& Astronomy, Monash University, Clayton, VIC 3800, Australia\\
$^{4}$Research School of Astronomy and Astrophysics, Australian National University, Canberra, ACT 2611, Australia
}

\pubyear{2016}

\begin{document}
\label{firstpage}
\pagerange{\pageref{firstpage}--\pageref{lastpage}}

\maketitle

\begin{abstract}
We perform a comparison between the smoothed particle magnetohydrodynamics (SPMHD) code, {\sc Phantom}, and the Eulerian grid-based code, {\sc Flash}, on the small-scale turbulent dynamo in driven, Mach 10 turbulence. We show, for the first time, that the exponential growth and saturation of an initially weak magnetic field via the small-scale dynamo can be successfully reproduced with SPMHD. The two codes agree on the behaviour of the magnetic energy spectra, the saturation level of magnetic energy, and the distribution of magnetic field strengths during the growth and saturation phases. The main difference is that the dynamo growth rate, and its dependence on resolution, differs between the codes, caused by differences in the numerical dissipation and shock capturing schemes leading to differences in the effective Prandtl number in {\sc Phantom} and {\sc Flash}.
\end{abstract}


\section{Introduction}

Supersonic turbulence regulates star formation \citep{mk04,mo07, hf12, padoanetal14}, producing the dense filaments that permeate molecular clouds along which dense cores and protostars form \citep[e.g.,][]{larson81, hartmann02, es04, hatchelletal05, andreetal10, perettoetal12, hacaretal16a, federrath16, kainulainenetal16}. That molecular clouds are magnetised cannot be ignored. Magnetic fields are no longer thought to prevent gravitational collapse altogether, but may still determine the rate and efficiency of star formation, even with weak magnetic fields, via super-Alfv\'enic turbulence \citep{nl08, lunttilaetal09, pb08, pb09, pn11, fk12, federrath15}. Therefore, it is vital to understand processes which determine the magnetic field strength inside molecular clouds. These processes involve highly non-linear dynamics making analytic study difficult (though see \citealt{sg94,gs95}), while observations of magnetic fields in molecular clouds are time consuming and only yield field directions in the plane of the sky and magnitudes along the line of sight \citep[e.g.,][]{crutcher99,bourkeetal01,ht05,tc08, crutcheretal10, crutcher12}. Numerical simulations can complement analytics and observations, and it is important to compare results from different codes to establish the conditions under which those results are representative of the physical processes involved. In this work, we focus on the small-scale turbulent dynamo as a mechanism for magnetic field amplification in molecular clouds, comparing calculations using smoothed particle magnetohydrodynamics (SPMHD) with those using grid-based methods.

\subsection{Small-scale turbulent dynamo}

The small-scale dynamo grows magnetic fields in a turbulent environment by the conversion of kinetic energy into magnetic energy. Operating near the dissipation scale, it is there that the smallest motions can efficiently grow the magnetic field through rapid winding and twisting of the magnetic field lines, with the magnetic energy growing exponentially via a reverse cascade of energy from small to large scales \citep[see review by][]{bs05}. 

The exponential growth rate is determined primarily by the physical viscosity and magnetic resistivity of the plasma, which can be expressed as dimensionless numbers: the kinematic Reynolds number (${\rm Re}=VL/\nu$, where $V$ is the velocity, $L$ is the characteristic length, and $\nu$ is the kinetic dissipation), the magnetic Reynolds number (${\rm Rm}=VL/\eta$, where $\eta$ is the resistive dissipation), and the ratio of the two, the `magnetic Prandtl number', ${\rm Pm}={\rm Rm}/{\rm Re}$ \citep{schekochihinetal04a, bs05, schoberetal12a, schoberetal12b, bss13, federrathetal14}. Using a large set of numerical simulations, \citet{federrathetal11} found that the dynamo growth rate is also dependent upon the compressibility of the plasma, parameterised by the turbulent Mach number, and is more efficient for turbulence driven by solenoidal (rotational) flows rather than compression. 

The magnetic field will saturate first at the dissipation scale, after which the dynamo enters a slow linear or quadratic magnetic energy growth phase \citep{choetal09, schleicheretal13}. This occurs due to the back-reaction of the Lorentz force on the turbulent flow as it begins to resist the winding of the magnetic field \citep{schekochihinetal02, schoberetal15}. The magnetic field on larger spatial scales will continue to slowly grow through reverse cascade of magnetic energy. Thus the small-scale dynamo can be considered as progressing through three distinct phases: i) the exponential growth phase, ii) the slow linear or quadratic growth phase once the magnetic energy is saturated at the dissipation scale, and iii) the saturation phase once the magnetic field has saturated on all spatial scales. 

\subsection{Previous turbulence comparisons}

Simulating magnetised, supersonic turbulence is challenging due to the range of flow conditions present and the need to uphold the divergence-free constraint of the magnetic field. Comparing results between different numerical methods is the best way to be confident in their results, and there have been several major code comparison projects related to supersonic turbulence. \citet{taskeretal08} compared two grid codes ({\sc Enzo}, {\sc Flash}) and two smoothed particle hydrodynamics (SPH) codes ({\sc Gadget}2, {\sc Hydra}) on simple test problems involving strong hydrodynamic shocks, finding comparable results when the number of particles were roughly equal to the number of grid cells. \citet{kitsionasetal09} studied decaying, supersonic, hydrodynamic (non-magnetised) turbulence, comparing four grid codes ({\sc Enzo}, {\sc Flash}, {\sc TVD}, {\sc Zeus}) and three SPH codes ({\sc Gadget}, {\sc Phantom}, {\sc Vine}). They found similar velocity power spectra and density probability distribution functions (PDFs) when the number of resolution elements were comparable, though the particle codes were found to be more dissipative. \citet{kritsuketal11} compared decaying, supersonic turbulence with magnetohydrodynamics (MHD) using nine different grid codes: {\sc Enzo}, {\sc Flash}, {\sc KT-MHD}, {\sc LL-MHD}, {\sc Pluto}, {\sc PPML}, {\sc Ramses}, {\sc Stagger}, and {\sc Zeus}. They found that all methods produced physically consistent results, with the quality of results improved with higher-order numerical solvers, and by exactly rather than approximately maintaining the divergence-free constraint on the magnetic field. 

A key limitation of both the \citet{kitsionasetal09} and \citet{kritsuketal11} comparisons is that they studied decaying turbulence. Interpolating the initial conditions obtained by driving the turbulence in one code introduced discrepancies between codes before the numerical experiments even started. Those discrepancies in the initial conditions were most severe between grid and particle methods, but also for different grid discretisations (e.g. staggered vs. unstaggered meshes), and is problematic in the MHD case since one must enforce $\nabla \cdot \bm{B}=0$. Furthermore, it is difficult to obtain a statistically significant sample of simulation snapshots in the absence of a statistical steady-state, given that supersonic turbulence decays within a few crossing times. This limitation means that intermittent, intrinsic fluctuations of the turbulence largely exceeded systematic differences in the numerical schemes, which we want to quantify.
 
\citet{pf10} (hereafter PF10) addressed these issues in a hydrodynamic comparison by using driven instead of decaying turbulence, allowing time-averaged statistical comparisons. They compared two codes: the grid code, {\sc Flash}, and the SPH code, {\sc Phantom}. The calculations started from a well-defined initial state of uniform density gas at rest, and both codes used exactly the same turbulence driving routine and force sequence to prevent any bias from different implementations. They found similar resolution requirements to previous studies, but that grid-based methods were better at resolving volumetric statistics at a given resolution, while SPH better sampled density-weighted quantities. However, this comparison was limited to hydrodynamic turbulence.

Recently, \citet{tp12} developed a new divergence cleaning method for SPMHD that maintains $\nabla \cdot \bm{B} = 0$ to sufficient accuracy for a wide range of problems, such as the simulation of jets and outflows during protostar formation \citep*{ptb12, btp14, lbp15, wpb16}. The best prior approach to maintain the divergence-free constraint was the Euler potentials, defining $\bm{B} = \nabla \alpha \times \nabla \beta$ \citep{pb07,rp07}. However, this excludes dynamo processes by construction because the Euler potentials cannot represent and follow wound-up magnetic field structures \citep{pb08, brandenburg10, price12}. By directly evolving the magnetic field and enforcing the divergence-free constraint with the constrained hyperbolic divergence cleaning method \citep{tp12}, it is now possible to simulate magnetohydrodynamic turbulence and MHD dynamos with SPMHD. Furthermore, \citet{tp13} improved the magnetic shock-capturing algorithm, particularly when dealing with weak magnetic fields in strong shocks.
 
\subsection{Outline}

This paper presents a code comparison on the small-scale dynamo amplification of a weak magnetic field from driven, supersonic turbulence. For our comparison, we use the same hydrodynamic codes, driving routine and Mach number as the \citetalias{pf10} comparison, so that any differences arise only from the MHD implementation. We investigate the three phases of small-scale dynamo amplification: exponential growth of magnetic energy, slow linear or quadratic growth as the magnetic field approaches saturation, and the fully saturated phase of the magnetic field.  In order to capture these phases completely, the calculations are evolved for one hundred crossing times, in contrast to only ten in \citetalias{pf10}. This is also intended to allow for high quality time-averaged statistics to be obtained so that a meaningful comparison can be made.

The paper is structured as follows: Section~\ref{sec:turbdetails} describes the details of the comparison. Results of the calculations are analysed in Section~\ref{sec:turbresults} and summarised in Section~\ref{sec:turbsummary}.

\section{Comparison details}
\label{sec:turbdetails}

We compare the codes {\sc Phantom} (SPMHD) and {\sc Flash} (grid). Both solve the ideal MHD equations but with fundamentally different numerical approaches: {\sc Flash} discretises all fluid variables into fixed grid points, whereas {\sc Phantom} discretises the mass of the fluid into a set of Lagrangian particles that move with the fluid velocity. The calculations are performed for a series of resolutions, using $128^3$, $256^3$, and $512^3$ resolution elements (grid points and particles, respectively).

\subsection{Magnetohydrodynamics}
\label{sec:mhd}

We solve the ideal MHD equations, namely
\begin{gather}
\frac{{\rm d}\rho}{{\rm d}t} = - \rho \nabla \cdot \bm{v} , \label{eq:continuity} \\
\frac{{\rm d}\bm{v}}{{\rm d}t} = - \frac{1}{\rho} \nabla \left( P + \frac{B^2}{2 \mu_0} \right) + \frac{1}{\mu_0 \rho} \nabla \cdot \left( \bm{B} \bm{B} \right) , \label{eq:momentum} \\
\frac{{\rm d}\bm{B}}{{\rm d}t} = (\bm{B}\cdot \nabla) \bm{v} - \bm{B} (\nabla \cdot \bm{v}), \label{eq:induction}
\end{gather}
where ${\rm d}/{\rm d}t = \partial/\partial t + \bm{v} \cdot \nabla$ is the material derivative, $\rho$ is the density, $\bm{v}$ is the velocity, $P$ is the pressure, $\bm{B}$ is the magnetic field, and $\mu_0$ is the permeability of free space.
The continuum equations have zero viscous and resistive dissipation (hence ideal). Since the growth rate of magnetic energy by small-scale dynamo amplification is set by dissipation, the growth rate in these calculations is set by the numerical dissipation present in the schemes. In Eulerian grid-based methods, numerical dissipation is introduced by the discretisation of the advection term in the material derivative. By contrast, Lagrangian particle-based methods compute the material derivative exactly. The shock-capturing scheme is the other primary source of numerical dissipation. Modern grid-based methods use Riemann solvers, introducing numerical dissipation related to the accuracy of the shock reconstruction. The approach in particle methods is to explicitly add artificial viscous and resistive terms in order to capture shocks, using switches to tune the dissipation to the relevant discontinuity.

\subsection{Initial conditions}
\label{sec:initcond}

The initial state is chosen to be simple so that both codes start from conditions which are identical. The initial density field is uniform with $\rho_0=1$, the velocity field is zero throughout ($\bm{v}=0$), and the calculation is performed in a periodic box of length $L = 1$. An isothermal equation of state, $P~=~c_{\rm s}^2 \rho$, is used to calculate the pressure with sound speed $c_{\rm s} = 1$. The magnetic field is set to $\sqrt{2}\times 10^{-5}$ in the $z$~-~direction. With $\mu_0=1$, this yields an initial plasma beta, the ratio of thermal to magnetic pressure, of $\beta=P/P_{\rm mag}=10^{10}$. This initial magnetic field is certainly not representative of actual magnetic fields in molecular clouds, but is intentionally chosen to be weak to observe the exponential growth of magnetic energy via the small-scale dynamo.

\subsection{Turbulent driving routine}
\label{sec:turbdriving}

As in \citetalias{pf10}, supersonic turbulence was initiated and sustained at a root mean square (rms) Mach number of $\mathcal{M} = 10$ by an imposed driving force generated from an Ornstein-Uhlenbeck process \citep{ep88, schmidtetal09, federrathetal10}. This is a stochastic process with a finite autocorrelation timescale. The driving force can be decomposed in Fourier space into longitudinal and solenoidal modes. We assume purely solenoidal driving, so that the turbulence is driven primarily by vorticity rather than compression (c.f. \citealt{federrathetal11, federrath13}). However, one third of the kinetic energy will still be contained in compressive modes due to the high Mach number of the turbulence \citep{ps10, federrathetal10}. 

Consistency of the driving pattern between codes was achieved by pre-generating the time-sequence of the Ornstein-Uhlenbeck modes, with both codes reading the pattern from file. The acceleration from the driving pattern is reconstructed at each individual cell or particle location by direct summation over all Fourier modes. The driving is at large scales, with a parabolic weighting of modes between $k_{\rm min} = 1$ and $k_{\rm max} = 3$, with smaller structures forming through turbulent cascade. The autocorrelation timescale is $1t_{\rm c}$, with $t_{\rm c}$ as defined in Equation~\ref{eq:timescale}. The driving routine was developed by \citet{federrathetal10}, and both the driving routine and the pattern file used in the paper are publicly available\footnote{The turbulent driving routine is bundled as part of the {\sc Flash} code (http://flash.uchicago.edu/site/flashcode/), and the pattern file used for these simulations is available at http://cita.utoronto.ca/$\sim$ttricco/mhdturbulence/.} with the input parameters used to generate the pattern file specified in Table~\ref{tbl:stir}. The stirring energy is used to obtain the variance of the Ornstein-Uhlenbeck process, corresponding to the autocorrelation time and energy input rate.

\begin{table}
\caption{Input parameters for the turbulence driving routine.}
\label{tbl:stir}
\centering
\begin{tabular}{ll}
\hline
Parameter & Value \\
\hline
spectral form & 1 (Parabola) \\
solenoidal weight & 1 \\
stirring energy & 8.0 \\
autocorrelation time & 0.05 \\
minimum wavenumber & 6.28 \\
maximum wavenumber & 18.90 \\
original random seed & 1 \\
\hline
\end{tabular}
\end{table}

The relevant physical timescale is the turbulent crossing time, which we define according to
\begin{equation}
t_{\rm c} \equiv \frac{L}{2 \mathcal{M} c_{\rm s}},
\label{eq:timescale}
\end{equation}
corresponding to $t_{\rm c} = 0.05$ in code units. The turbulence was simulated for 100 crossing times, covering the full growth phase of the dynamo up until the magnetic energy reaches its saturation level, with at least half of the total time spent in the saturation phase.  

\subsection{{\sc Phantom} -- SPMHD}
\label{sec:phantom}

{\sc Phantom} is a smoothed particle magnetohydrodynamics (SPMHD) code. The MHD equations (Equations \ref{eq:continuity}--\ref{eq:induction}) are implemented as described in \citet{pm04a, pm04b, pm05} and \citet{price12}, using \citet*{bot01}'s method of subtracting $\bm{B} (\nabla \cdot \bm{B})$ from the momentum equation to keep the magnetic tensional force stable. This implementation of momentum and induction equations resolves issues related to non-zero $\nabla \cdot \bm{B}$ in a manner that is equivalent to the Powell 8-wave approach \citep{powell94, powelletal99}. In addition to this, we employ the constrained hyperbolic divergence cleaning method of \citet{tp12}, which is an SPMHD adaptation and improvement of the cleaning algorithm by \citet{dedneretal02}. The cleaning wave speed is set to the local fast MHD wave speed. During the course of this work, it was found that if the wave speed included the term involving the relative velocity of particles (as in the artificial viscosity), then the individual timestepping scheme could introduce significant errors to the magnetic field. This occurred when particles were interacting on timestep bins that were spaced too far apart. Using a timestep limiter \citep[i.e.,][]{sm09} can prevent these errors, but for these calculations we simply reduce the cleaning speed by excluding the relative velocity.

Shocks are captured by adding an artificial viscosity, as described by \citet{pm04a, pm05} and based on the \citet{monaghan97} formulation. It is important that the signal velocity, defining the characteristic speed of information propagation, includes a term involving the relative motion of particles to prevent particle interpenetration \citep{monaghan89}, and as found by \citet{pf10}, it is necessary to increase this by setting the dimensionless constant to $\beta_{\rm AV}=4$ for Mach 10 shocks (as opposed to the common $\beta_{\rm AV}=2$). We use the \citet{mm97} switch to reduce dissipation away from shocks.

Discontinuities in the magnetic field are treated with an artificial resistivity \citep{pm04a, pm05}. {\sc Phantom} uses a new switch we have recently developed to reduce dissipation of the magnetic field away from discontinuities \citep{tp13}. This switch solves problems with the switch proposed by \citet{pm05}, namely that it is able to capture shocks when the sound speed is significantly higher than the Alfv\'en speed (i.e., in the super-Alfv\'enic regime when the magnetic field is very weak).  This is done by using the dimensionless quantity $h\vert \nabla \bm{B} \vert / \vert \bm{B} \vert$, which measures the relative strength of the discontinuity in the magnetic field. 

The smoothing length (resolution length), $h$, of each particle is calculated in the usual manner by iteration of the density summation with $h = 1.2 (m / \rho)^{1/3}$ using a Newton-Raphson solver \citep{pm04b, pm07}. This means that the numerical resolution scales with the density. For these set of simulations, the resolution increases by $4$--$8\times$ in the highest density regions, with a decrease in resolution of about $2\times$ in the lowest density regions.  Timesteps are set individual to each particle in a scheme that is block hierarchical in powers of two, with each particle setting its timestep based on its local Courant condition.  Second order leapfrog time integration is used.

\subsection{{\sc Flash} -- Grid code}
\label{sec:flash}

{\sc Flash}\footnote{http://flash.uchicago.edu/site/flashcode} is a grid-based code using a finite volume scheme for solving the MHD equations \citep{fryxelletal00, dubeyetal08}. Although {\sc Flash} can be used with adaptive mesh refinement \citep[AMR,][]{bc89}, our simulations employ a fixed and uniform cartesian grid for simplicity. We here use {\sc Flash} with the HLL3R approximate Riemann solver for ideal MHD, based on a MUSCL-Hancock scheme \citep*{wfk11}. This is a predictor-corrector scheme and is second-order accurate in both space and time. \citet{wfk11} further show that this MHD scheme maintains $\nabla \cdot \bm{B}\sim0$ to within negligible errors, by using divergence cleaning in the form of the parabolic cleaning method of \citet{marder87} \citep[see also][]{dedneretal02}. The MHD solver is particularly efficient and robust because it uses a relaxation technique that guarantees positive density and gas pressure and thus avoids unphysical states, by construction.

\subsection{Computational cost}
\label{sec:compcost}

%
%
%

%
%
%

The {\sc Flash} calculations used $90$, $1600$, and $40~000$ cpu-hours for the $128^3$, $256^3$, and $512^3$ simulations. The {\sc Phantom} calculations used $2700$ and $44~000$ cpu-hours for the $128^3$ and $256^3$ simulations, and $280~000$ cpu-hours for the $512^3$ calculation from $t=0\to40t_{\rm c}$. It is expected that each factor of 2 increase in resolution should increase the computational expense by $16\times$, since there are $8\times$ more resolution elements and the Courant condition should reduce the timestep by half. Both codes exhibit a scaling behaviour that is close to this. For {\sc Phantom}, the particles are spread over $\sim6$, $7$, and $8$ individual timestep bins for the $128^3$, $256^3$, and $512^3$ resolution calculations, respectively.  Approximately $35\%$ of the computational expense in the {\sc Phantom} calculations is spent on neighbour finding. The driving routine adds negligible computational expense ($\sim2\%$ of overall cpu-hours).

{\sc Flash} uses distributed memory parallelisation via MPI (message passing interface), with the $128^3$, $256^3$, and $512^3$ resolution calculations performed on 8, 64, and 512 cores, respectively. At the time these calculations were performed, the MPI implementation in {\sc Phantom} was not finalised, so the {\sc Phantom} calculations used shared memory parallelisation via OpenMP, using 40 cores on a single node for all calculations. The {\sc Flash} calculations required 11, 25, and 80 wall clock hours, respectively, representing modest increases with resolution even though the cpu-hour cost increased by $20$--$25\times$ for each factor two increase in resolution. The $128^3$ resolution {\sc Phantom} calculation required $68$ wall clock hours to be run to completion, but $1060$ wall clock hours ($\sim 6$ weeks) for the $256^3$ calculation. The difference in wall clock times between the two codes results from the differing parallelisation methods, since {\sc Phantom}, unlike {\sc Flash}, could not use additional cores for the higher resolution calculations in order to reduce the wall clock time. Overall, we find that, at comparable resolution, the {\sc Phantom} calculations take roughly $30\times$ more cpu-hours than the {\sc Flash} calculations, and that, as in \citetalias{pf10}, the $256^3$ {\sc Phantom} calculation takes approximately an equivalent amount of computational time as the $512^3$ {\sc Flash} calculation.

\subsection{Analysis methods}
\label{sec:analysismethods}

\subsubsection{Power spectra}

Power spectra are calculated using the same analysis tool for both codes to ensure that results are comparable. The {\sc Flash} data is directly analysed with this tool, while the power spectrum of {\sc Phantom} data is obtained by interpolating the particles to a grid of double the particle resolution (i.e., $256^3$ particles are interpolated to $512^3$ grid points). A higher resolution grid is chosen in order to represent the energy contained in the highest density structures, which are up to $4$--$8\times$ higher than the initial resolution. Appendix~\ref{sec:gridinterp} investigates the effect of the resolution of the interpolated grid, in addition to the difference between mass and volume weighted interpolation. We found that the magnetic field was satisfactorily represented by a grid which has twice the resolution of the particle calculation.

\subsubsection{Probability distribution functions}

Computing a volume-weighted PDF from grid methods involves binning the cells according to the value of the quantity and normalizing such that the integral under the PDF is unity. For SPH this is more complicated since the resolution is tied to the mass rather than the volume. \citetalias{pf10} computed the PDF directly from the SPH particles by weighting the contribution of each particle, $i$, by the volume element $m_{i}/\rho_{i}$. \citet*{pfb11} later found that this was inaccurate at high Mach number because $\sum_{i} m_{i}/\rho_{i}$ has no requirement that it equals the total volume. Instead, one should interpolate to a fixed volume using the SPH kernel $W$, since $m_{i}/\rho_{i}$ is only meaningful when multiplied by the kernel (since SPH is derived assuming $\sum_{i} m_{i}/\rho_{i} W(\bm{r} - \bm{r}_{i}, h) = 1$). However, interpolation to a fixed grid \citep[e.g.][]{kitsionasetal09} is also problematic since the resolution in our simulations is 4--8$\times$ higher in the densest regions compared to a fixed grid with the same number of resolution elements. Hence, sampling the high density tail of the SPH calculation would require a commensurably high resolution grid. We follow \citet{pfb11} in using an adaptive mesh to compute the PDF from the SPH particles, where the mesh is refined until the cell size is smaller than the smoothing length. The SPH PDF is then computed and normalised directly from this adaptive mesh.


\section{Results}
\label{sec:turbresults}

We focus of the analysis of our calculations on effects produced by the small-scale dynamo. Since this comparison uses the same codes, initial conditions, and turbulent driving routine as the hydrodynamic turbulence comparison of \citetalias{pf10}, analyses performed by \citetalias{pf10} have only been repeated where the addition of magnetic fields would be expected to alter the result (i.e., for the density PDF).

We analyse the three phases of small-scale dynamo amplification throughout this section: i) the steady, exponential growth of magnetic energy, ii) the slow linear or quadratic growth of magnetic energy once the magnetic field is saturated on the smallest scales, and iii) the fully saturated phase of magnetic energy. Since we assume ideal MHD, the numerical dissipation varies with resolution thus the kinetic and magnetic Reynolds numbers are not constant. This affects the growth rate and saturation level of magnetic energy, which enable us to contrast the scaling behaviour of the two methods.

\subsection{Formation of the turbulence; $t/t_{\rm c} \lesssim 2$}
\label{sec:transientphase}

\begin{figure}
\centering
\setlength{\tabcolsep}{0.004\linewidth}
 \begin{tabular}{ccc}
 \scriptsize{\sc $128^3$} & \scriptsize{\sc $256^3$} & \scriptsize{\sc $512^3$} \\
  \includegraphics[width=0.325\linewidth]{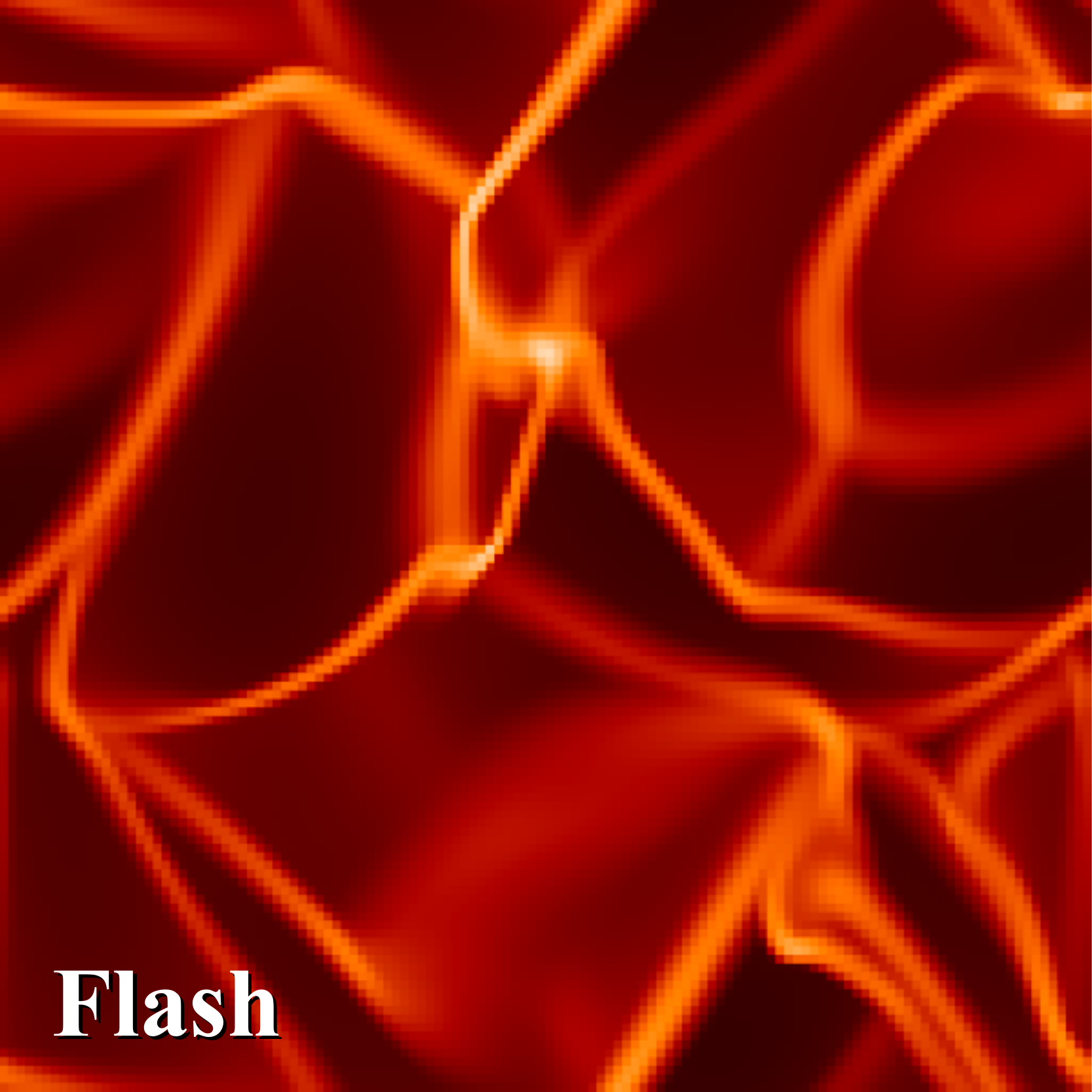} 
& \includegraphics[width=0.325\linewidth]{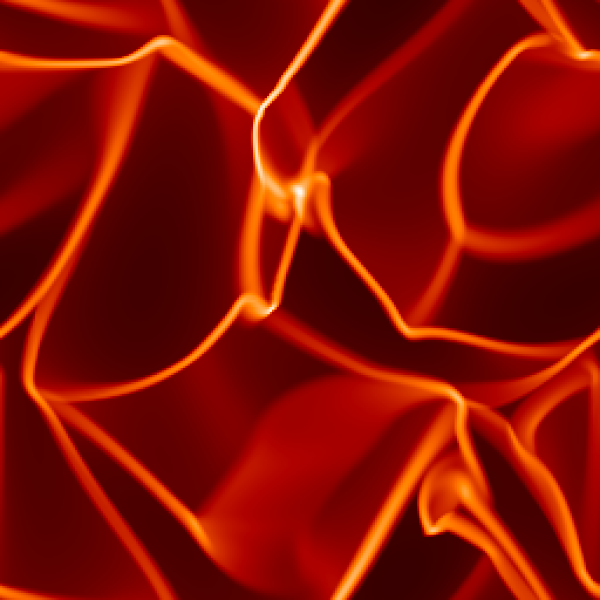} 
& \includegraphics[width=0.325\linewidth]{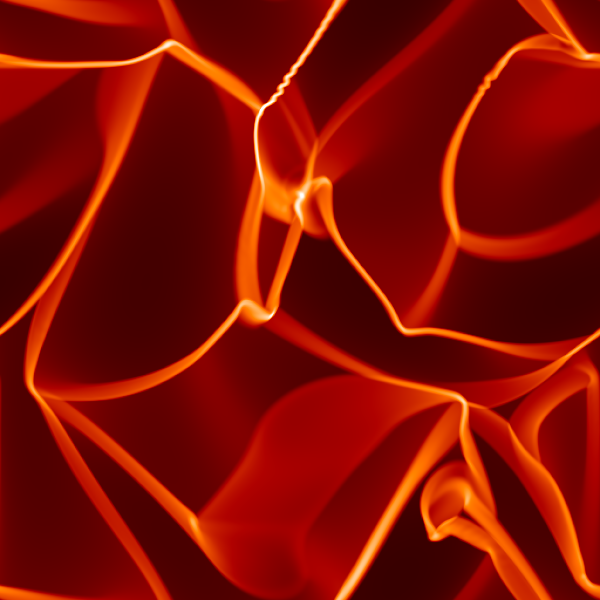} \\ 
  \includegraphics[width=0.325\linewidth]{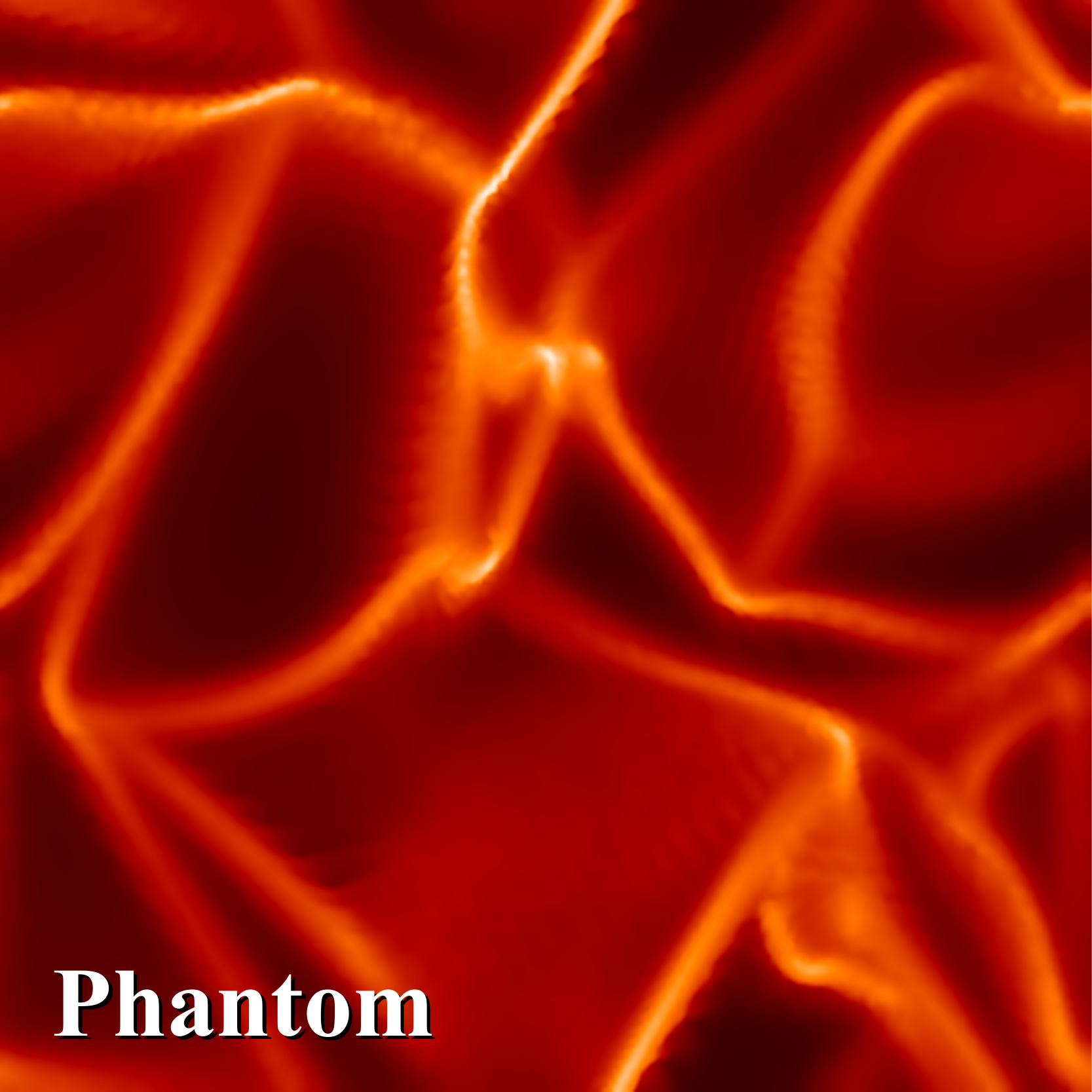} 
& \includegraphics[width=0.325\linewidth]{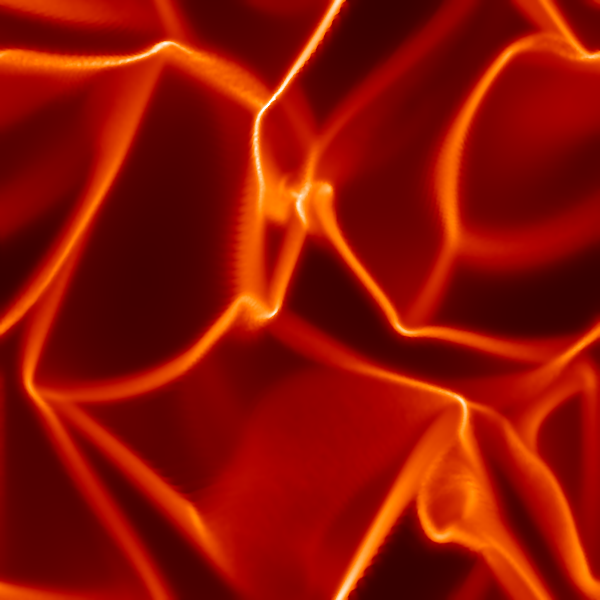} 
& \includegraphics[width=0.325\linewidth]{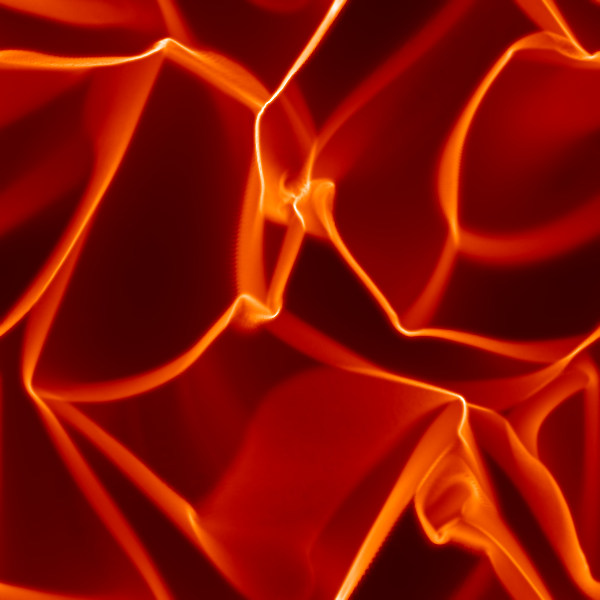} 
\end{tabular} \\
\includegraphics[width=\linewidth]{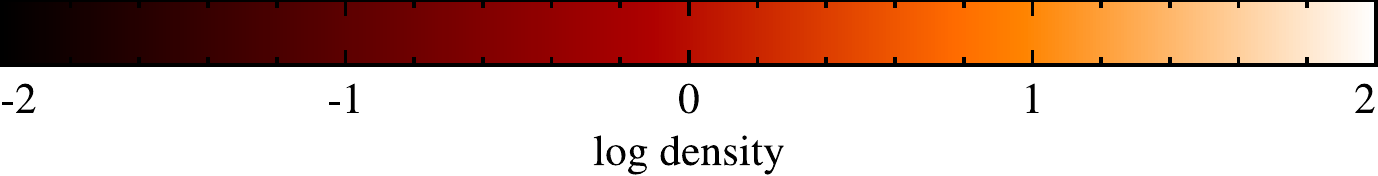}

 \begin{tabular}{ccc}
 \scriptsize{\sc $128^3$} & \scriptsize{\sc $256^3$} & \scriptsize{\sc $512^3$} \\
  \includegraphics[height=0.325\linewidth]{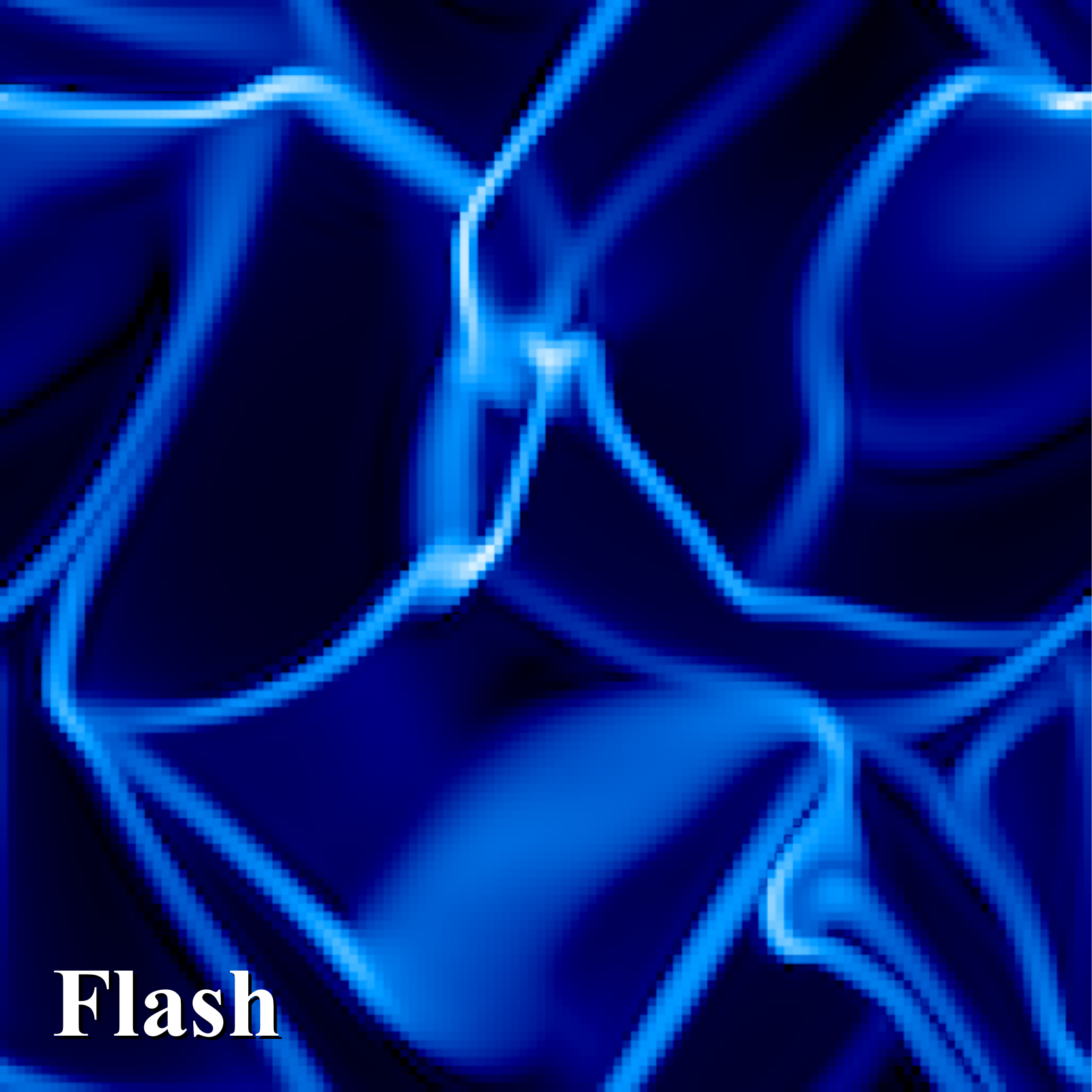} 
& \includegraphics[height=0.325\linewidth]{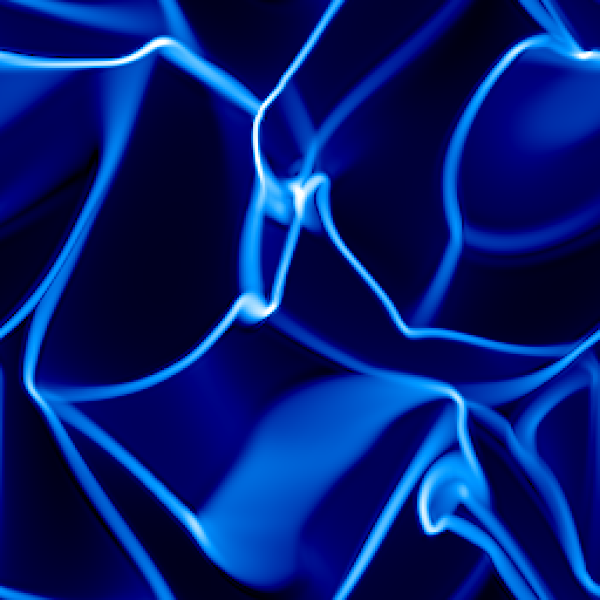} 
& \includegraphics[height=0.325\linewidth]{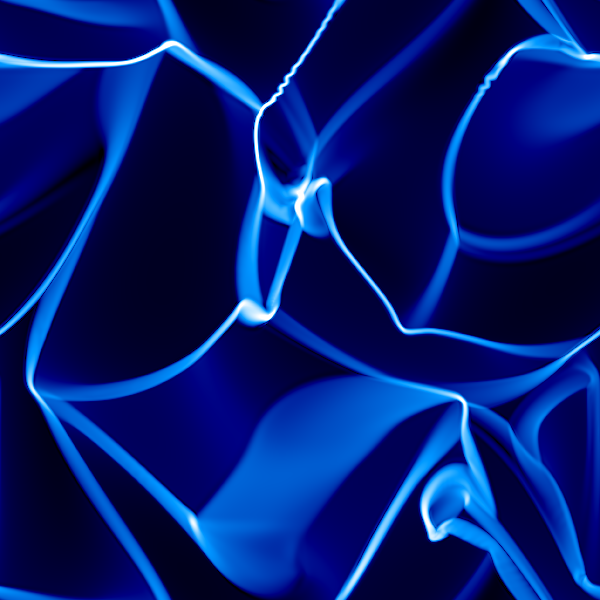} \\
  \includegraphics[height=0.325\linewidth]{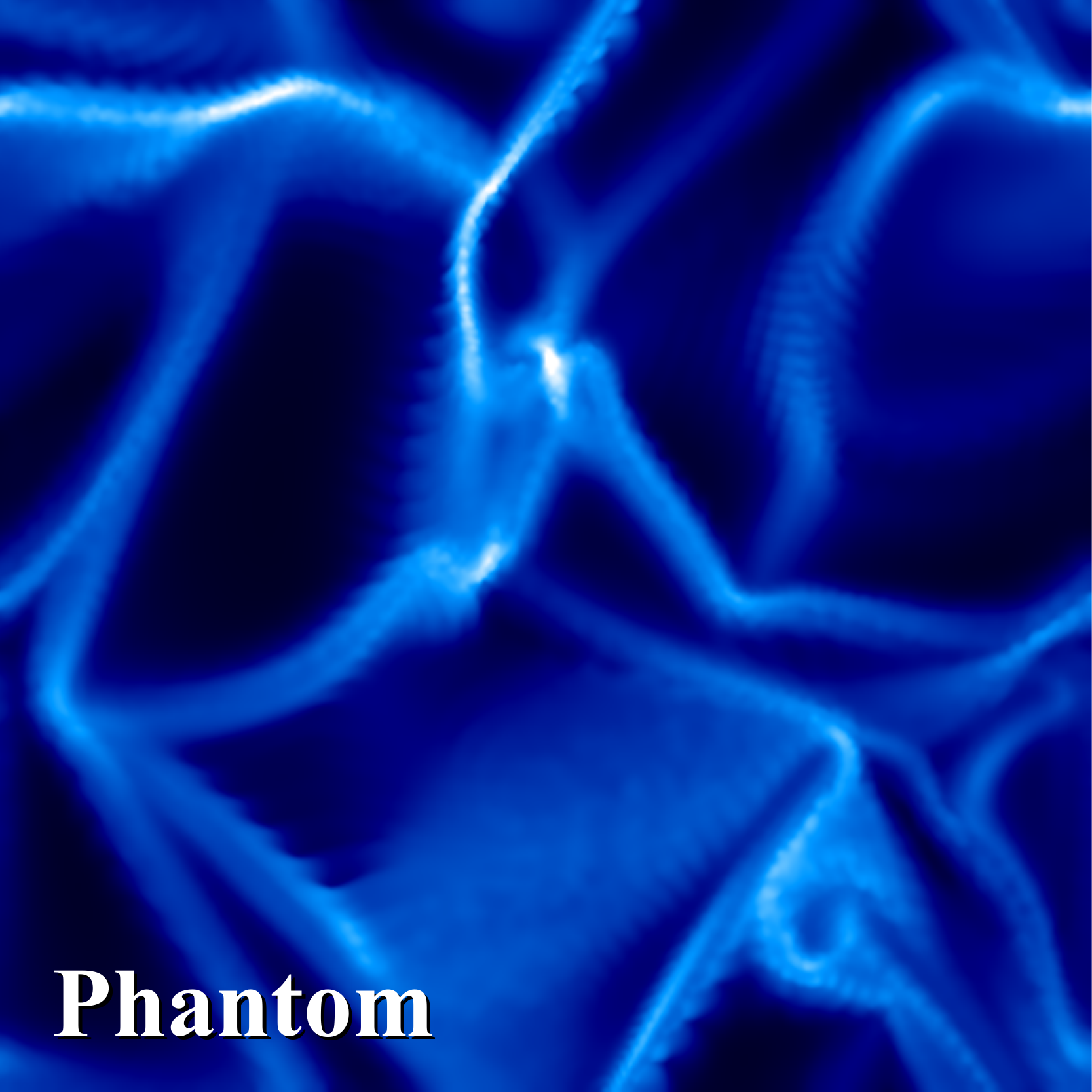} 
& \includegraphics[height=0.325\linewidth]{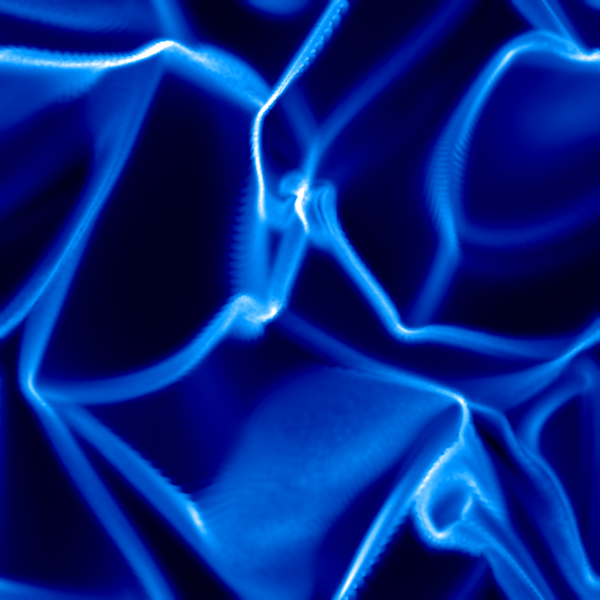} 
& \includegraphics[height=0.325\linewidth]{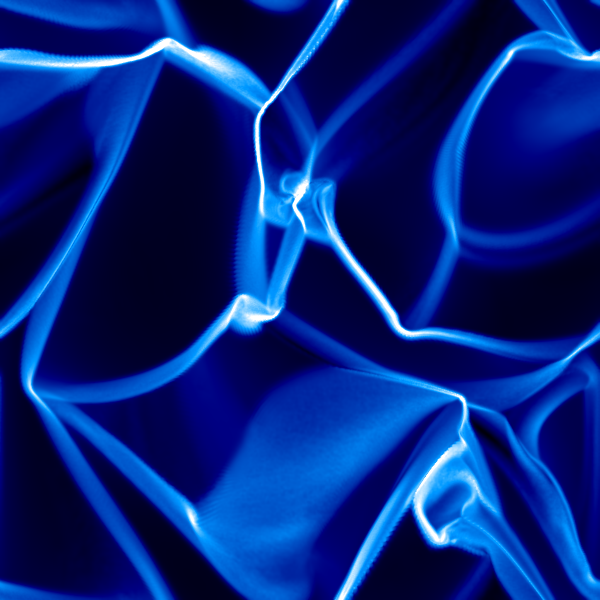} 
\end{tabular}
\includegraphics[width=\linewidth]{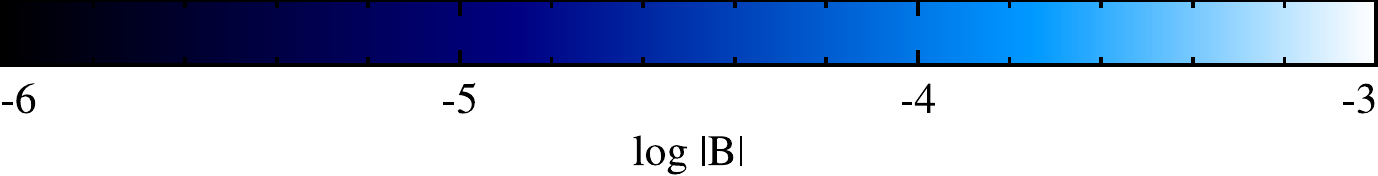}
\caption{Slices of $\rho$ (top) and $\vert B \vert$ (bottom) in the $z=0.5$ mid-plane at $t/t_{\rm c}=1$ during the initial formation of the turbulence. The results from {\sc Flash} (top row) and {\sc Phantom} (bottom row) are shown for resolutions of $128^3$, $256^3$, and $512^3$ (left to right). As the resolution is increased, the shock lines become more well defined. The regions with highest magnetic field strength are in the dense shocks.}
\label{fig:slices}
\end{figure}

The simulations begin with a brief transitory phase while the turbulence is formed by the driving routine. Fig.~\ref{fig:slices} shows slices of $\rho$ and $\vert B \vert$ at $z=0.5$ for $t/t_{\rm c}=1$, shortly after large shocks have been formed by the driving routine and started to interact. The magnetic field is strongest in regions where the density is highest due to compression of the magnetic field in the shocks. Conversely, the low density regions exhibit relatively weaker magnetic fields.

\begin{figure}
\centering
 \includegraphics[width=\linewidth]{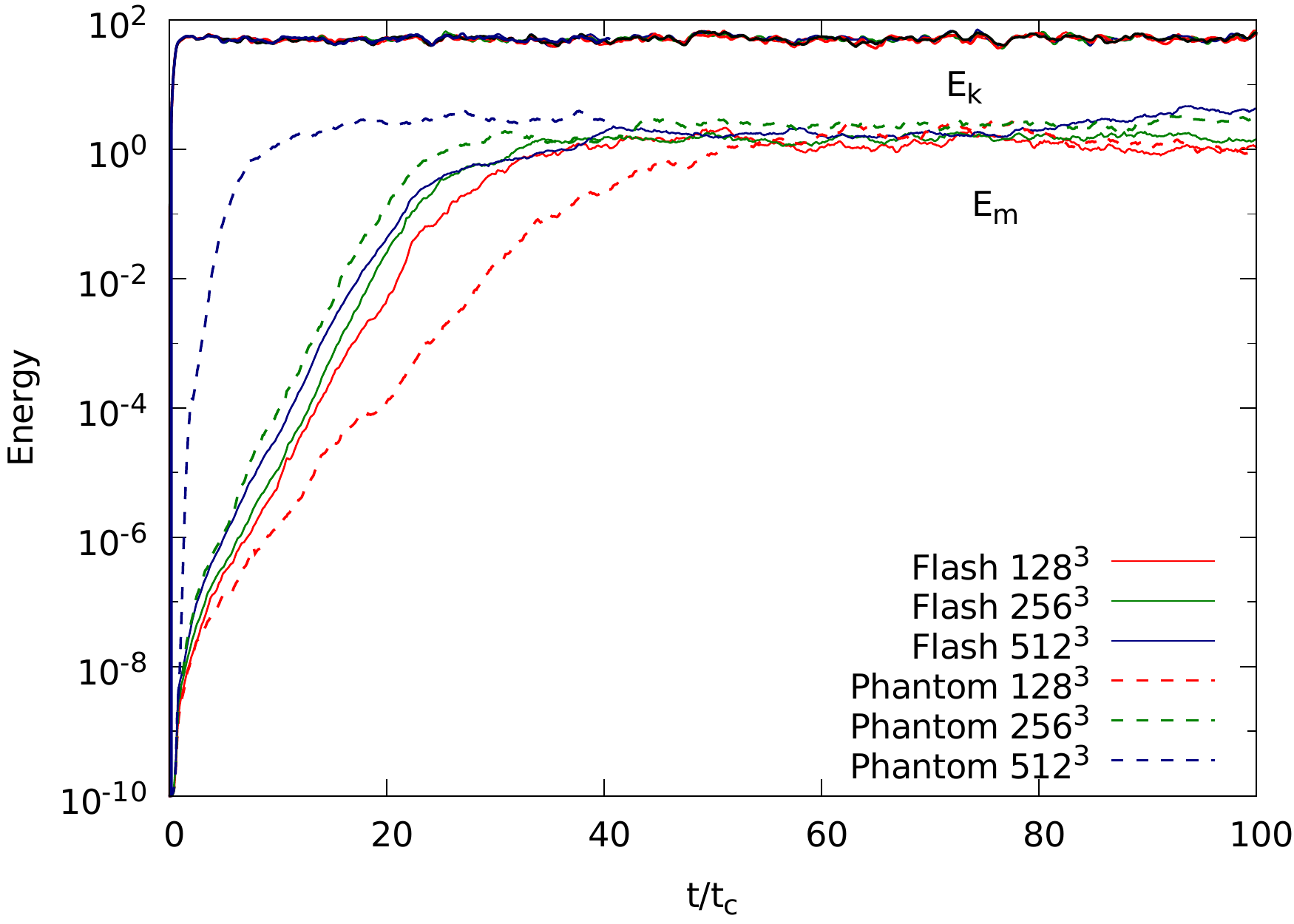}
\caption{Growth and saturation of the magnetic energy for {\sc Flash} and {\sc Phantom} at resolutions of $128^3$, $256^3$, and $512^3$ grid points and particles.  The steady, exponential growth of magnetic energy by the small-scale dynamo can be clearly seen in all calculations, along with the slow turnover of magnetic energy growth as the dynamo begins to saturate the magnetic energy and it enters the linear or quadratic growth phase. Both codes saturate the magnetic energy at similar levels. {\sc Flash} has similar magnetic energy growth rates across the resolutions simulated, while {\sc Phantom} exhibits faster growth rates with increasing resolution. This resolution dependence is a consequence of the artificial dissipation terms used for shock capturing. The top lines are the kinetic energy for the six calculations, kept at a steady level by the turbulent driving routine.}
\label{fig:en_mag}
\end{figure}

Approximately half a crossing time is required for the kinetic energy to saturate (see Fig.~\ref{fig:en_mag}), since the large-scale shocks contain the bulk of the kinetic energy, though it takes another turbulent crossing time before the turbulence is fully developed at smaller spatial scales. The magnetic energy is amplified by two orders of magnitude during the transient phase while the turbulence is developing (Fig.~\ref{fig:en_mag}). This occurs in two steps. First, for $t/t_{\rm c} \lesssim 1$, a sharp rise in magnetic energy is caused by the formation of large-scale shocks (Fig.~\ref{fig:slices}). Second, from $t/t_{\rm c} \approx 1$--2, the magnetic energy increases exponentially during the generation of small-scale structure in the density and magnetic fields caused by the interaction of the shocks, but at a rate higher by a factor of 2--3 than the measured rate in the exponential growth phase (Section~\ref{sec:growthrates}). Once the turbulence is fully developed on all spatial scales, the magnetic field enters the steady, exponential growth phase of the small-scale dynamo.

The initial transient growth of the magnetic field is resolution dependent, with higher resolutions resulting in higher magnetic energy by the time the turbulence is fully developed. For example, the magnetic energy in the $512^3$ {\sc Phantom} calculation is increased by an additional $3$--$4$ orders of magnitude compared to the other calculations. We have investigated whether this growth is merely a numerical artefact of the timestepping by re-doing the initial phase with a reduction in the Courant factor, and also by using global timesteps instead of individual timesteps. These did not alter our results. Additionally, we checked if this growth is driven by spurious generation of divergence of the magnetic field by both turning off the hyperbolic divergence cleaning (no divergence control), and conversely by increasing the hyperbolic cleaning wave speed by a factor of $10$ \citep[matching the rms velocity; see the over-cleaning method in][]{tricco15}. These showed the same fast transient magnetic field growth, so this is not caused by unphysical magnetic field growth in the form of high $\nabla \cdot \bm{B}$. Hence, the growth of magnetic energy in the {\sc Phantom} simulations appears to be physical, originating from the explicitly added dissipation terms rather than occurring due to numerical error or instability.

%

%
%

\subsection{Column integrated density and magnetic field strength}

\begin{figure*}
\centering
\setlength{\tabcolsep}{0.0025\columnwidth}
\renewcommand{\arraystretch}{1.0}
\begin{tabular}{ccccl}
 \multicolumn{4}{c}{\sc Flash} & \\
  \includegraphics[height=0.225\textwidth]{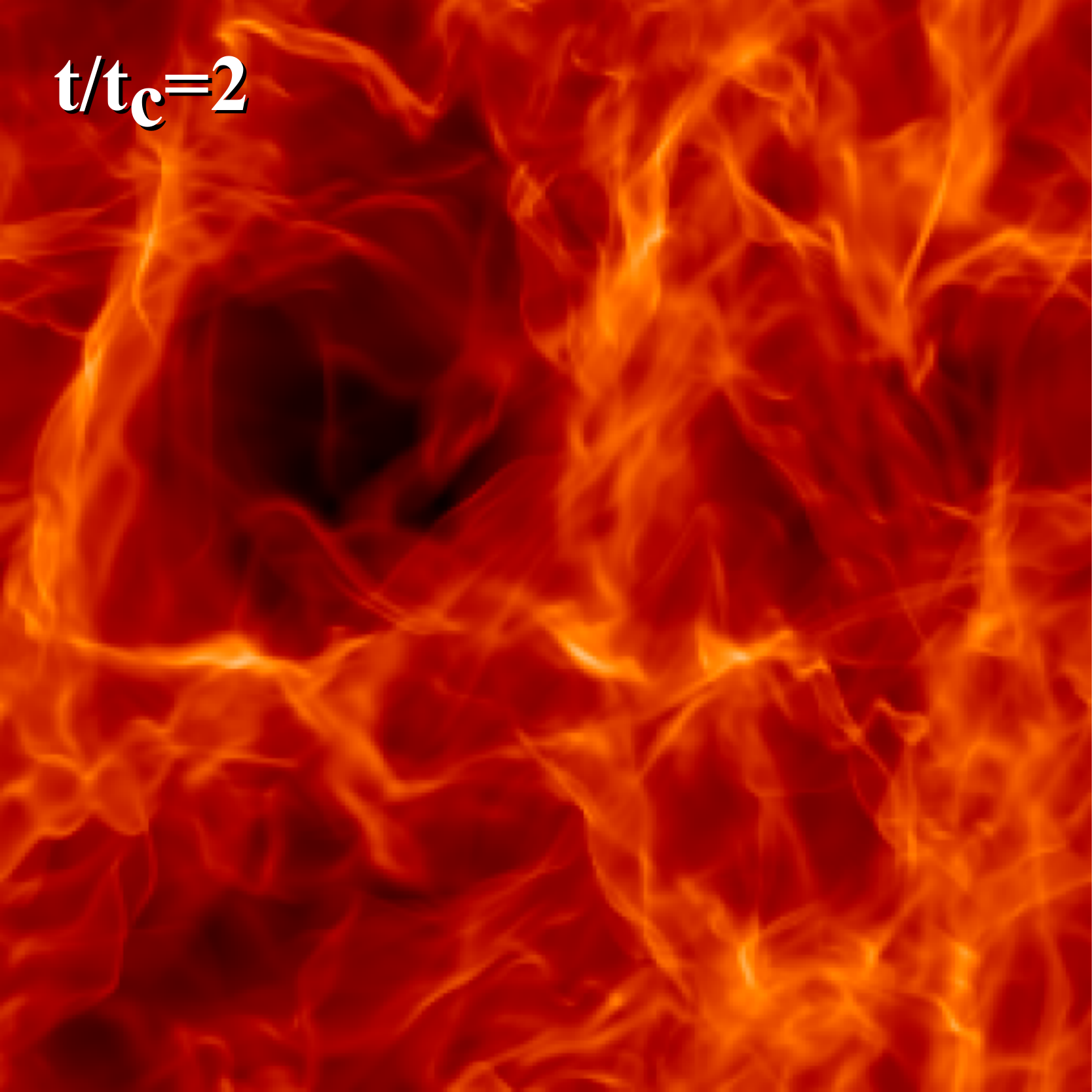}
& \includegraphics[height=0.225\textwidth]{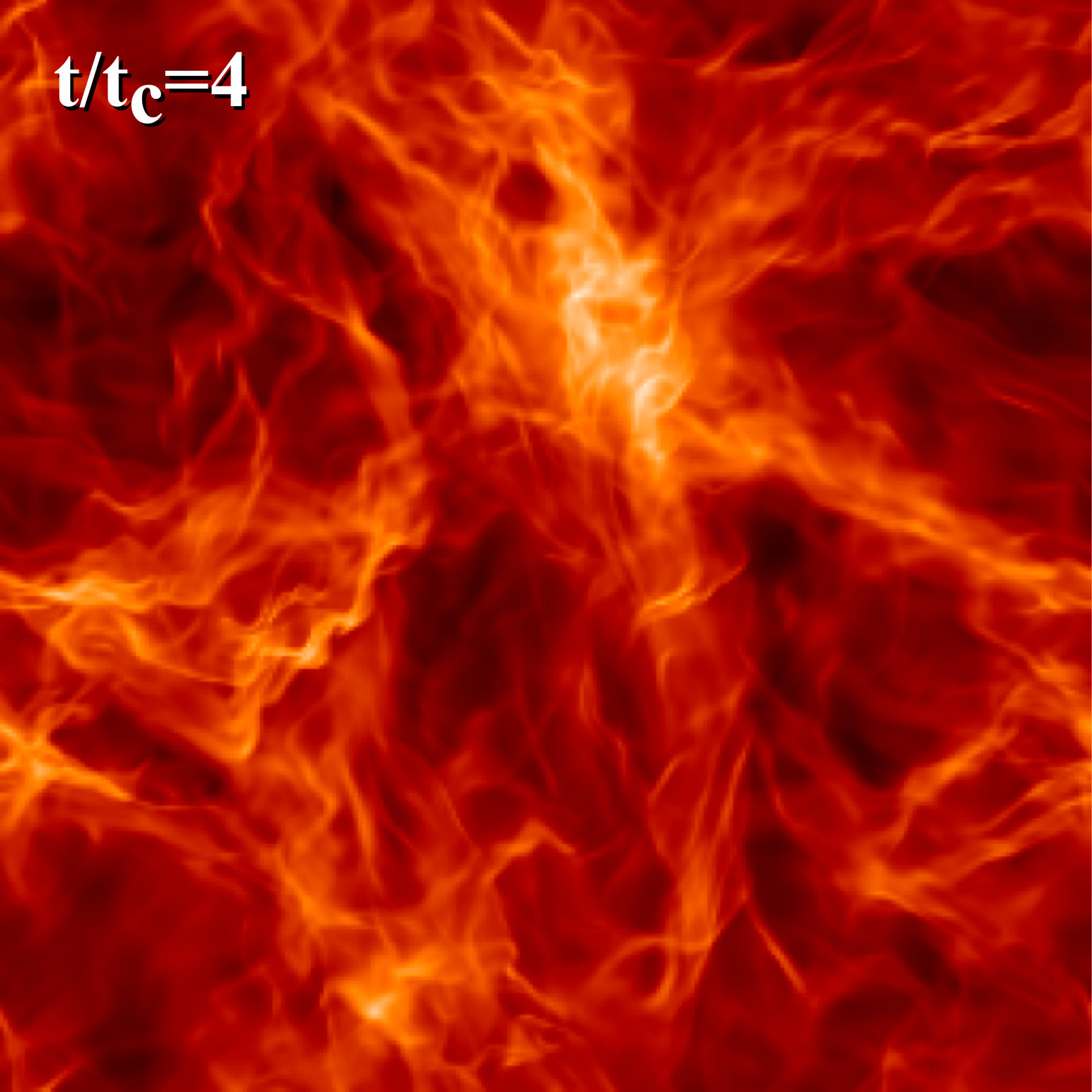}
& \includegraphics[height=0.225\textwidth]{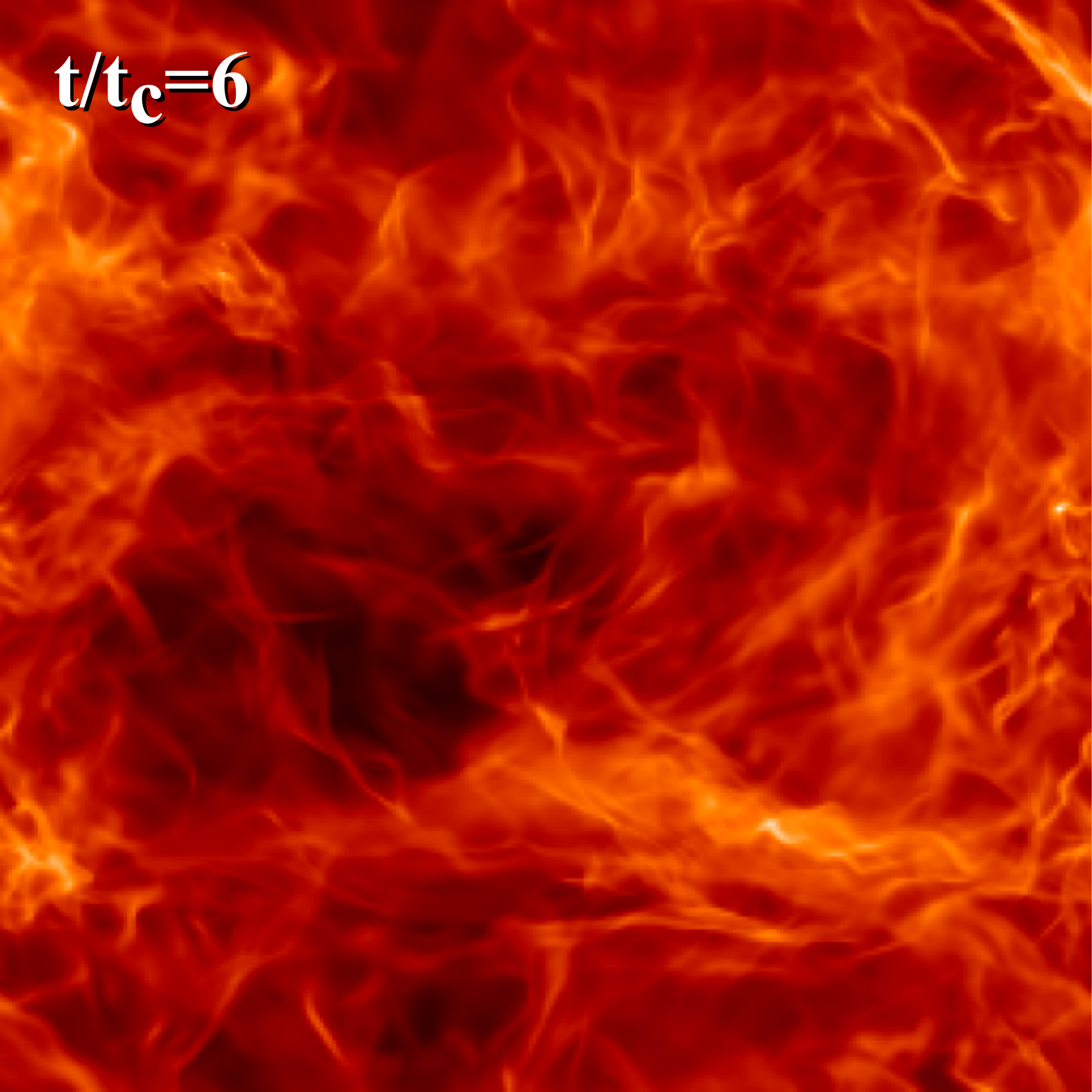}
& \includegraphics[height=0.225\textwidth]{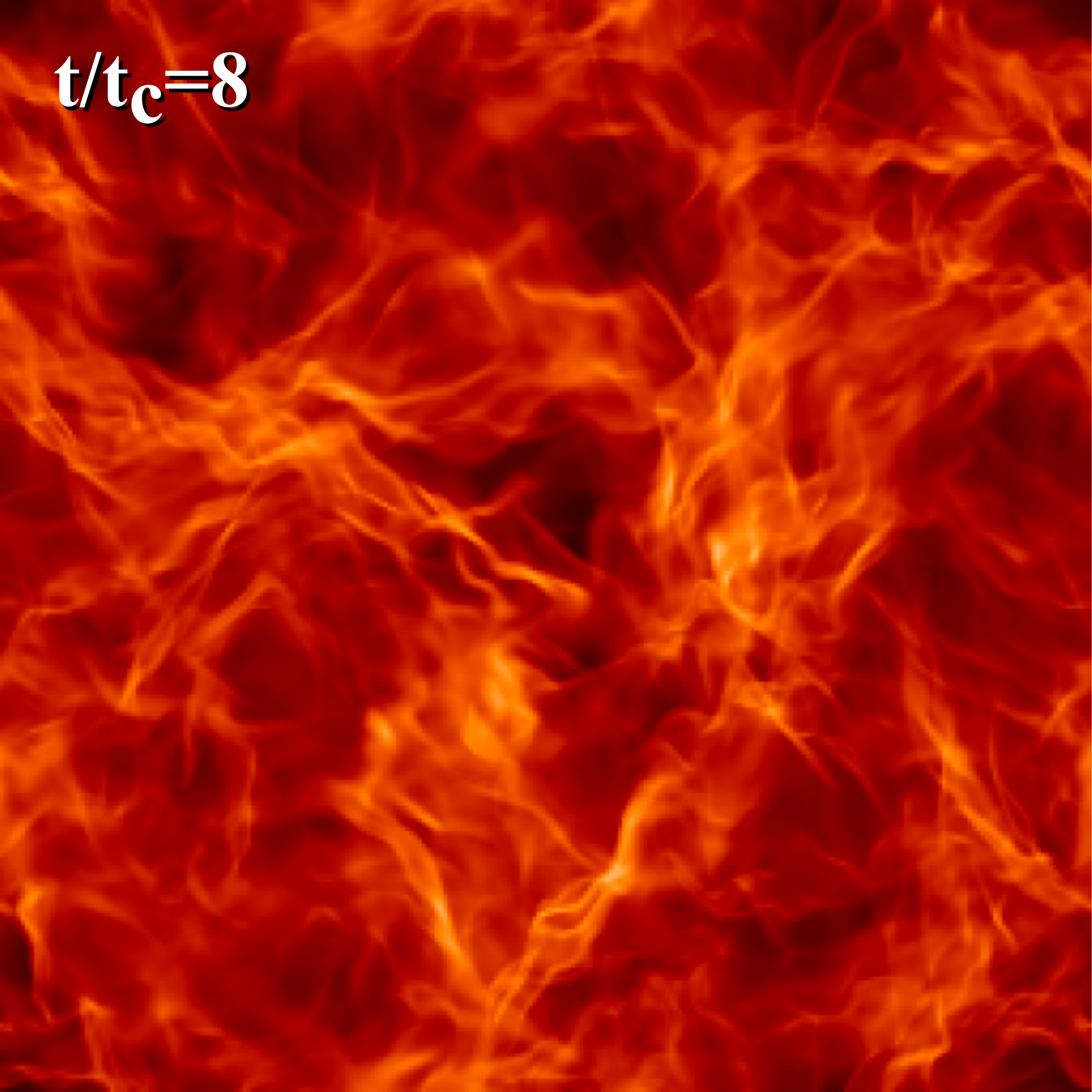} 
& \includegraphics[height=0.225\textwidth]{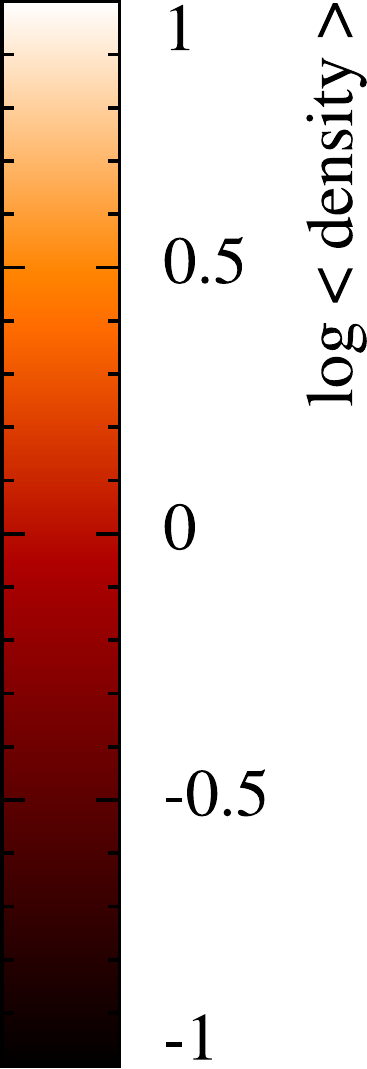} \\
  \includegraphics[height=0.225\textwidth]{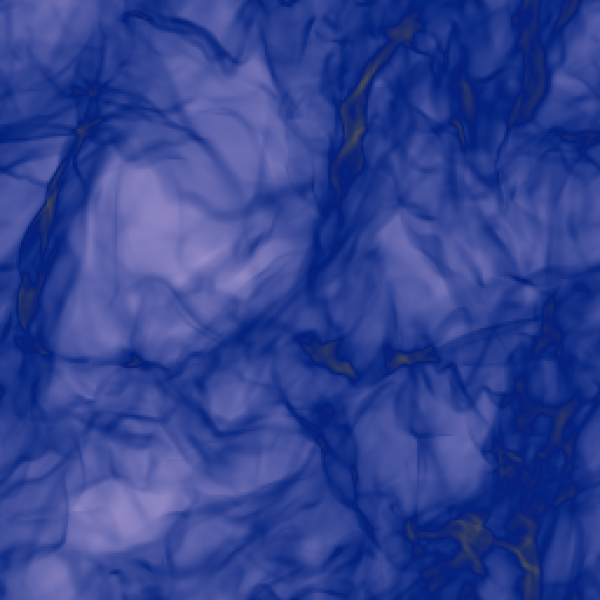}
& \includegraphics[height=0.225\textwidth]{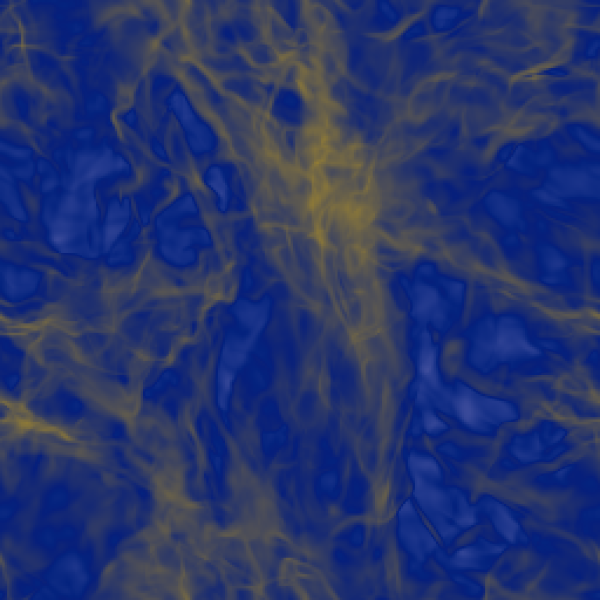}
& \includegraphics[height=0.225\textwidth]{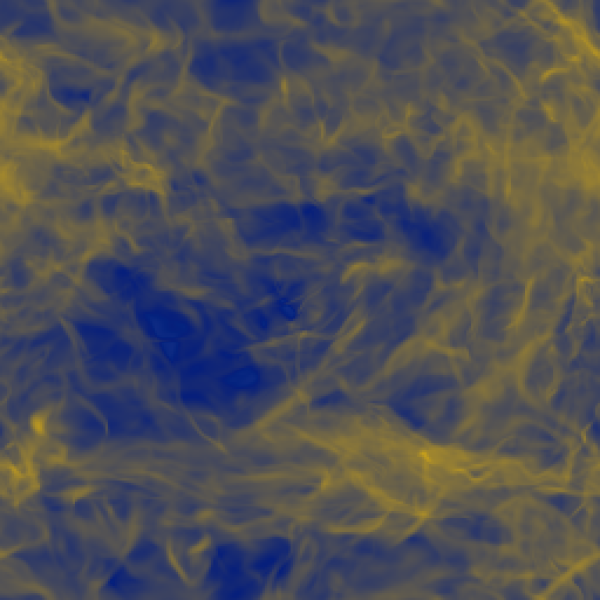}
& \includegraphics[height=0.225\textwidth]{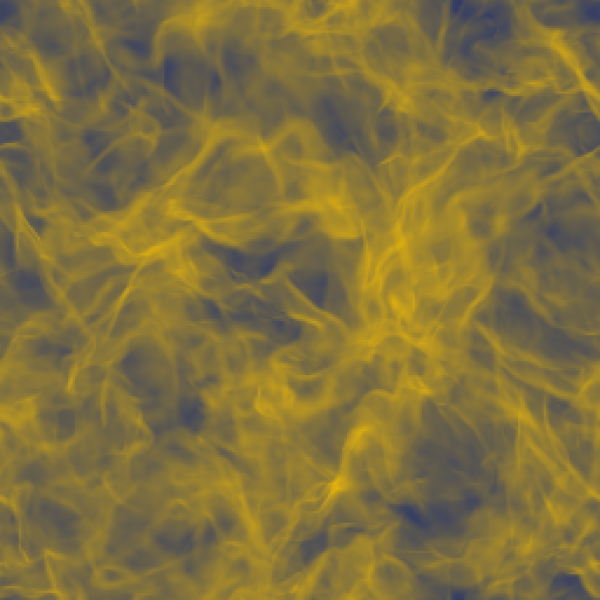}
& \includegraphics[height=0.225\textwidth]{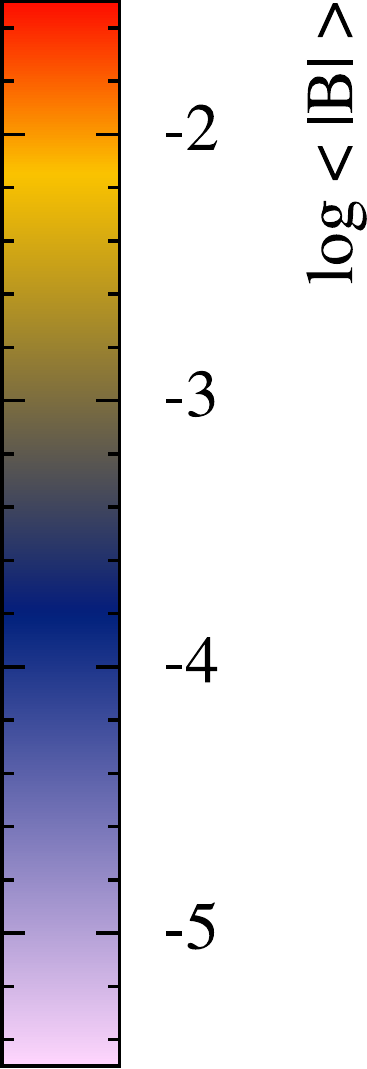}
\end{tabular} \\
\vspace{0.5cm}
\begin{tabular}{ccccl}
 \multicolumn{4}{c}{\sc Phantom} & \\
  \includegraphics[height=0.225\textwidth]{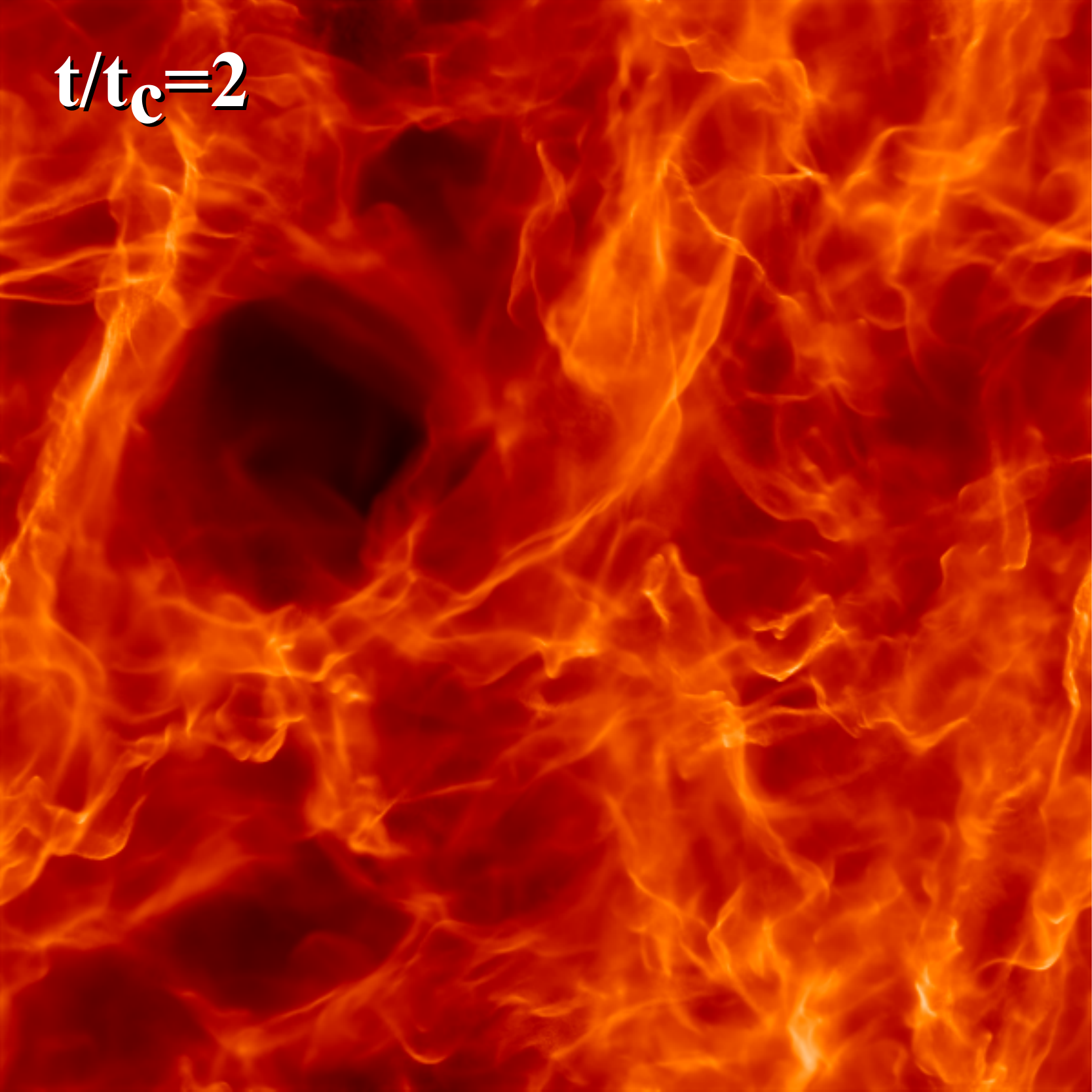}
& \includegraphics[height=0.225\textwidth]{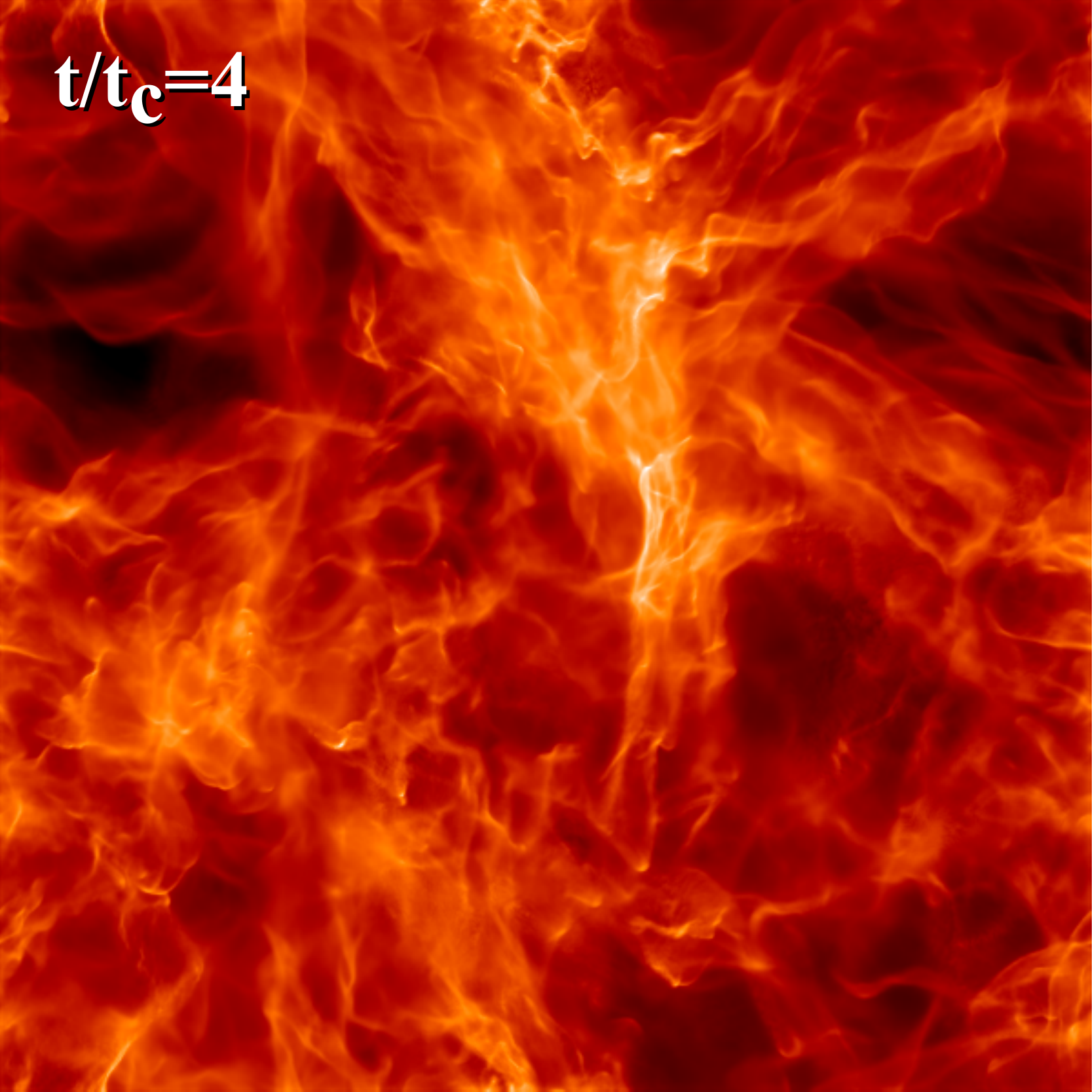}
& \includegraphics[height=0.225\textwidth]{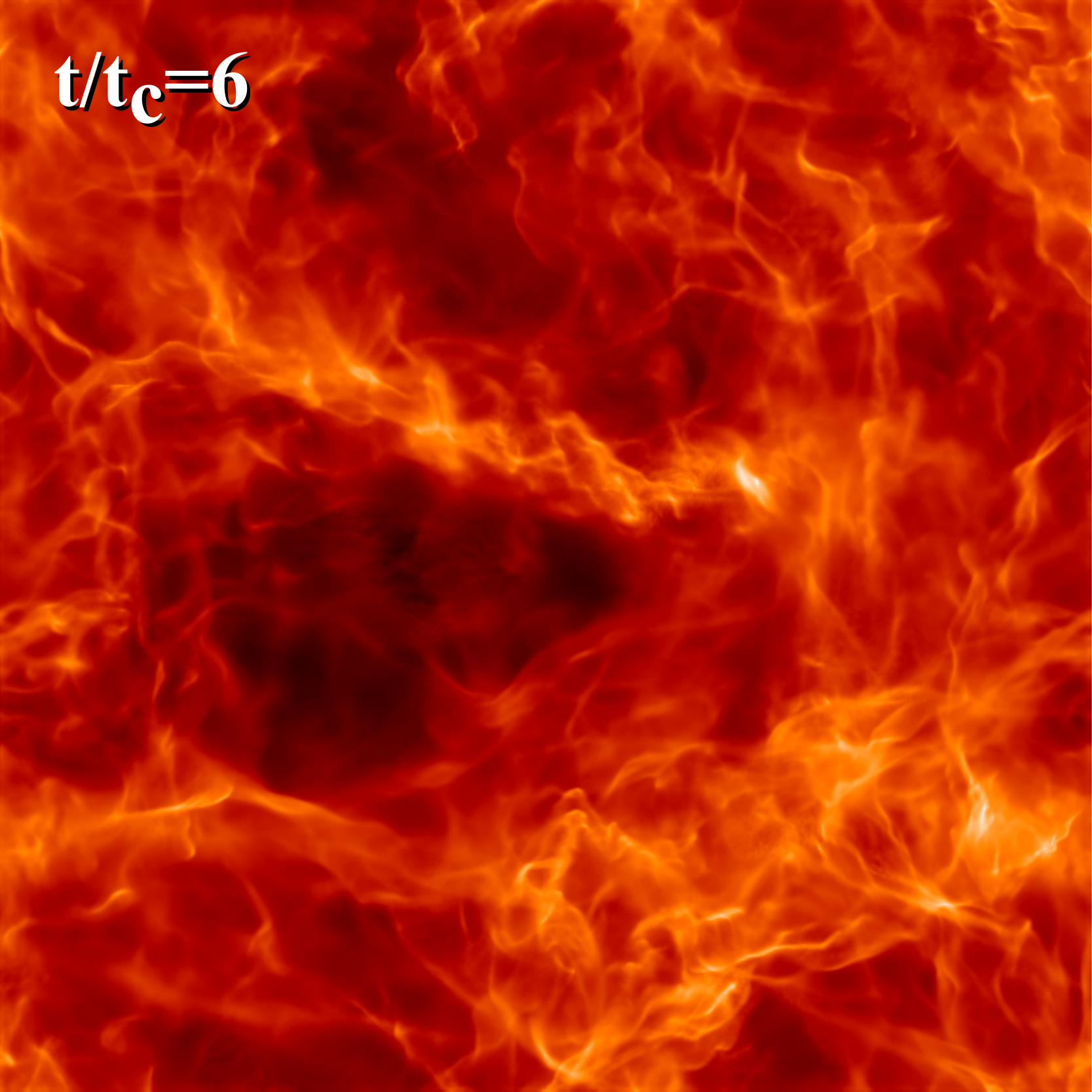}
& \includegraphics[height=0.225\textwidth]{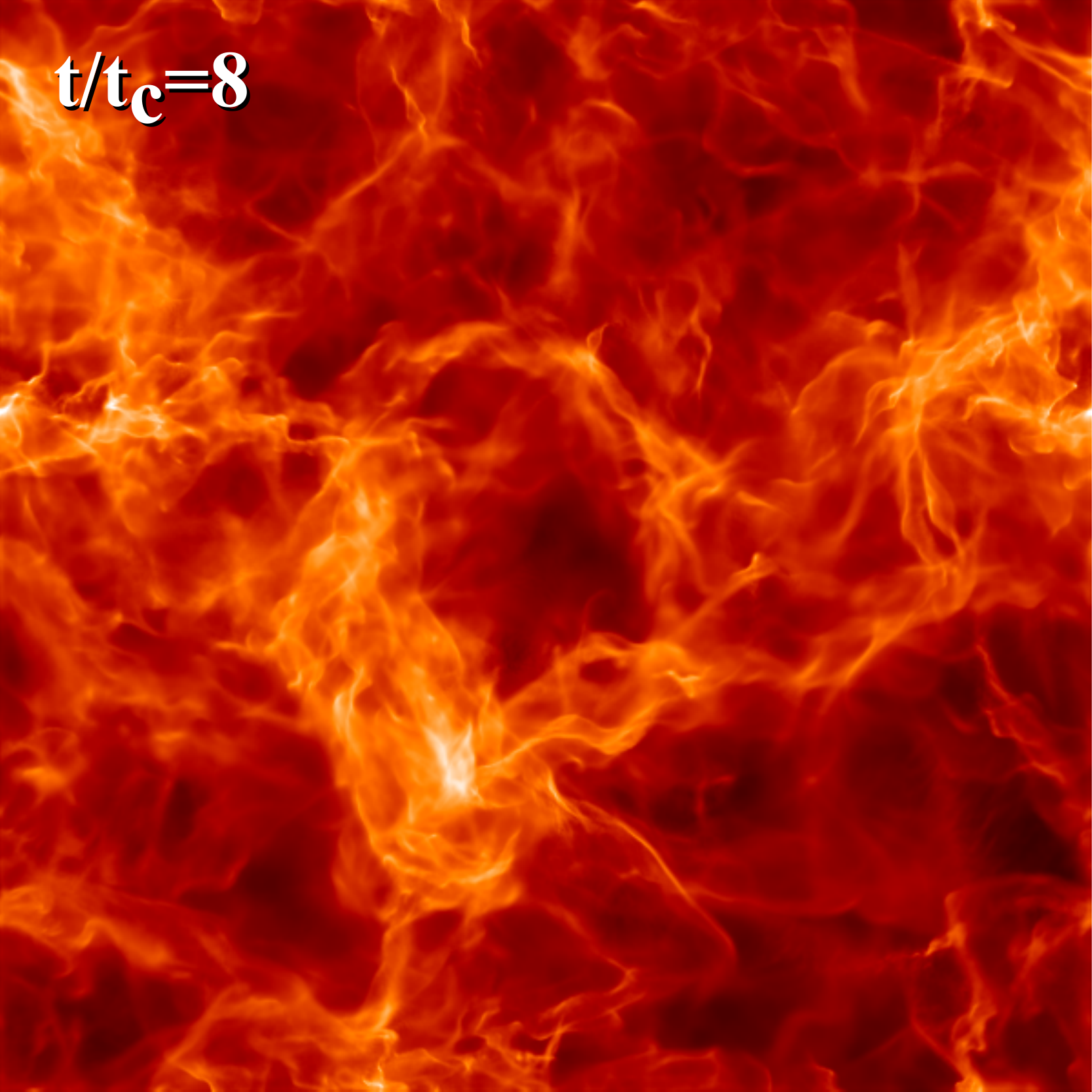}
& \includegraphics[height=0.225\textwidth]{cobar-column-rho-red.pdf} \\
  \includegraphics[height=0.225\textwidth]{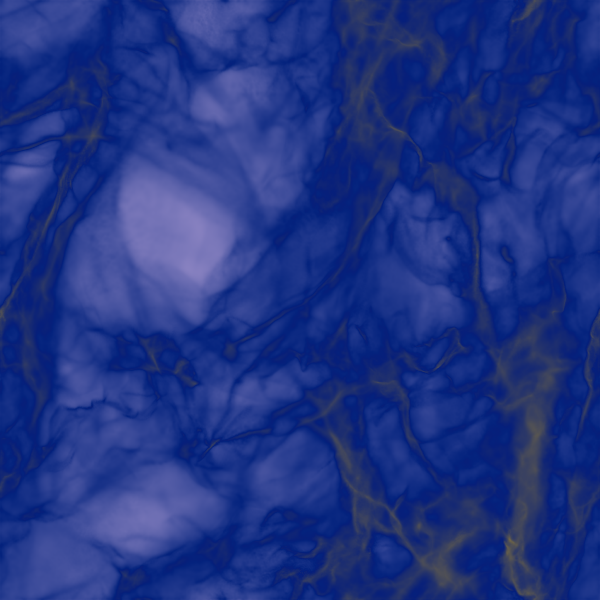}
& \includegraphics[height=0.225\textwidth]{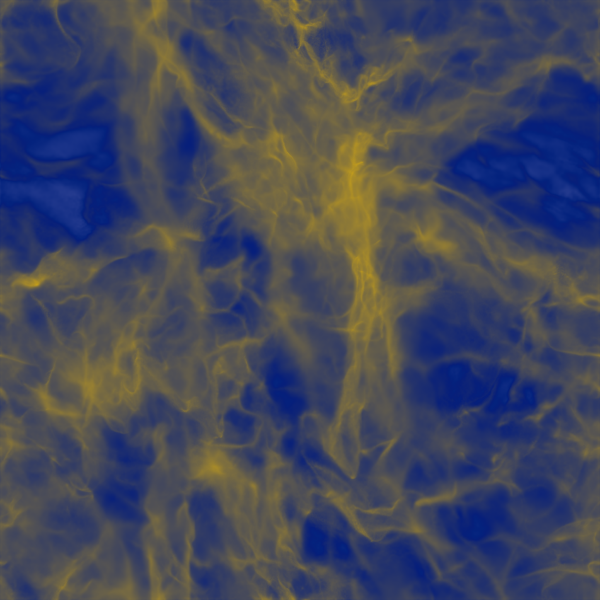}
& \includegraphics[height=0.225\textwidth]{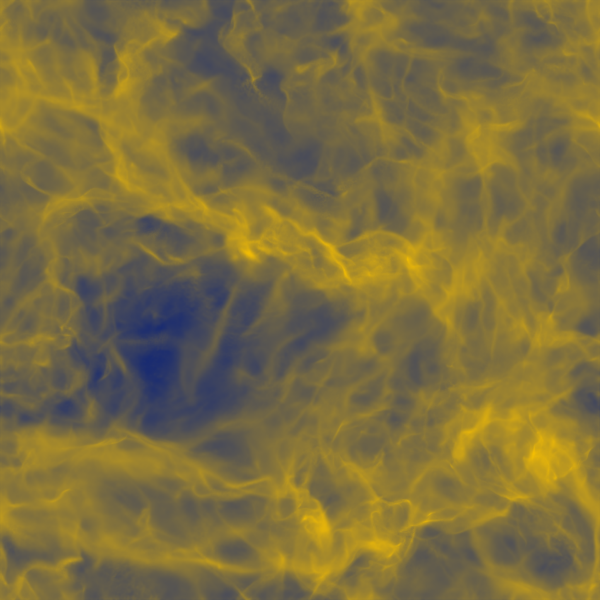} 
& \includegraphics[height=0.225\textwidth]{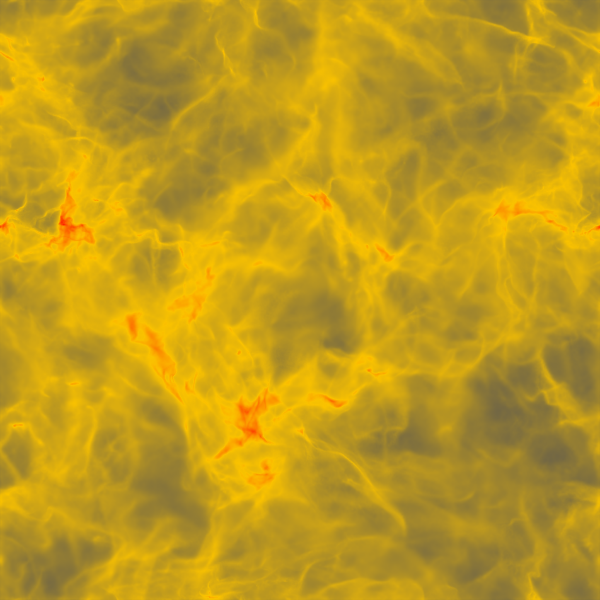} 
& \includegraphics[height=0.225\textwidth]{cobar-column-B-wbyr.pdf} 
\end{tabular}
\caption{$z$-column integrated $\rho$ and $\vert B \vert$, defined $<B> = \int \vert B \vert {\rm d}z / \int {\rm d}z$, for {\sc Flash} (top) and {\sc Phantom} (bottom) at resolutions of $256^3$ for $t/t_{\rm c}=2,4,6,8$. The density field has similar structure in both codes at early times, but diverge at late times due to the non-determinstic behaviour of the turbulence. The magnetic field is strongest in the densest regions, while the mean magnetic field strength throughout the domain increases with time.}
\label{fig:column-integrated}
\end{figure*}

Fig.~\ref{fig:column-integrated} shows a time sequence of column density and column integrated $\vert B \vert$ from $t/t_{\rm c}=2$--$8$, comparing {\sc Flash} (top figure) and {\sc Phantom} (bottom figure) calculations at $256^3$ since the growth rates are similar at this resolution (c.f. Fig.~\ref{fig:en_mag} and Table~\ref{tbl:energies}). Both codes show similar patterns in column density and magnetic field for the first few crossing times (left two columns), but eventually the patterns diverge due to the chaotic, non-deterministic nature of turbulence (right two columns) This was also found in \citetalias{pf10}.

There exists a definite correlation between the high density regions with the regions of strongest magnetic field when compared at a fixed time for each code individually. This is caused in part due to the compression of the gas, as similarly evidenced during the initial formation of the turbulence (Section~\ref{sec:transientphase}). Despite the high density regions having the highest magnetic field strength, the mean magnetic field strength throughout the domain can be seen to be increasing with time, a signature of the small-scale dynamo. This is quantitatively examined in Section~\ref{sec:magspectra} by computing the power spectrum of the magnetic energy.

\subsection{Exponential growth rate of magnetic energy}
\label{sec:growthrates}


\begin{table}
\caption{The growth rate ($\Gamma$) of magnetic energy during the exponential growth phase, defined $\propto \exp(\Gamma t / t_{\rm c})$, and the time-averaged saturation values of kinetic and magnetic energy.}
\label{tbl:energies}
\centering
\begin{tabular}{cccccc}
\hline
Calculation & $\Gamma$ & $\langle E_{\rm k} \rangle_\text{sat}$ & $\langle E_{\rm m} \rangle_\text{sat}$ \\ \hline
{\sc Flash} $128^3$ & 0.69 & 51.11 $\pm$ 5.51 & 1.20 $\pm$ 0.31  \\ 
{\sc Flash} $256^3$ & 0.75 & 51.19 $\pm$ 4.81 & 1.46 $\pm$ 0.20  \\ 
{\sc Flash} $512^3$ & 0.74 & 52.17 $\pm$ 5.15 & 2.36 $\pm$ 1.02  \\ 
{\sc Phantom} $ 128^3$ & 0.47 & 50.30 $\pm$ 4.80 & 1.52 $\pm$ 0.48  \\ 
{\sc Phantom} $ 256^3$ & 0.78 & 51.17 $\pm$ 5.34 & 2.31 $\pm$ 0.47  \\ 
{\sc Phantom} $ 512^3$ & 1.63 & 51.79 $\pm$ 3.94 & 2.98 $\pm$ 0.35  \\
\hline
\end{tabular}
\end{table}

The evolution of the magnetic energy as a function of time is shown in Fig.~\ref{fig:en_mag} for the calculations from both {\sc Phantom} and {\sc Flash} using $128^{3}$, $256^{3}$ and $512^{3}$ resolution elements (see legend). To compare the exponential growth rate of magnetic energy between the calculations, the magnetic energy in Fig.~\ref{fig:en_mag} during the exponential growth phase, defined between $t/t_{\rm c}=3$ and the onset of the slow growth phase, was fitted to $E_{\rm m} \propto \exp(\Gamma t/t_{\rm c})$ where $\Gamma$ is the growth rate. The growth rates are given in Table~\ref{tbl:energies}. The {\sc Flash} results all have similar growth rates. In contrast, the {\sc Phantom} results have growth rates that increase with resolution by nearly a factor of two for each doubling of resolution. 

Analytic studies of the exponential growth rate of the small-scale dynamo have shown that for $\text{Pm} \ll 1$, the growth rate scales with $\text{Rm}^{1/2}$, while for $\text{Pm} \gg 1$, it scales with $\text{Re}^{1/2}$ \citep{schoberetal12a, bss13}. Theoretical predictions of the growth rate for $\text{Pm}\sim 1$, which is the Prandtl number regime for numerical codes in the absence of explicit dissipation terms, are more uncertain.  The growth rate in the transition region between $0.1 < \text{Pm} < 10$ was probed by \citet{federrathetal14} using {\sc Flash} simulations with explicit viscous and resistive dissipation. They found that the magnetic energy growth rate for $\text{Pm} \lesssim 1$ exhibited a steep dependence on $\text{Pm}$ and only agreed qualitatively with the analytical expectations of \citet{schoberetal12a} and \citet{bss13}. Conversely, the growth rate for $\text{Pm} \gtrsim 1$ quantitatively agreed with analytical expectations, with, by comparison, relatively little variation with respect to $\text{Pm}$.

\citet{federrathetal11} measured the effective Prandtl number in {\sc Flash} through comparison with calculations with physical dissipation terms, finding that $\text{Pm}\sim2$. This is in agreement with similar experiments by \citet{lb07}. For {\sc Phantom}, the effective Prandtl number can be estimated analytically from the artificial dissipation terms, for which we find that $\text{Pm}\sim1$ for these calculations (see Appendix~\ref{sec:prandtl} for further discussion). Though the theory for the growth rate around $\text{Pm} \sim 1$ is still uncertain, our results appear to be in agreement with those of \citet{federrathetal14}.

\begin{figure}
 \centering
\includegraphics[width=\columnwidth]{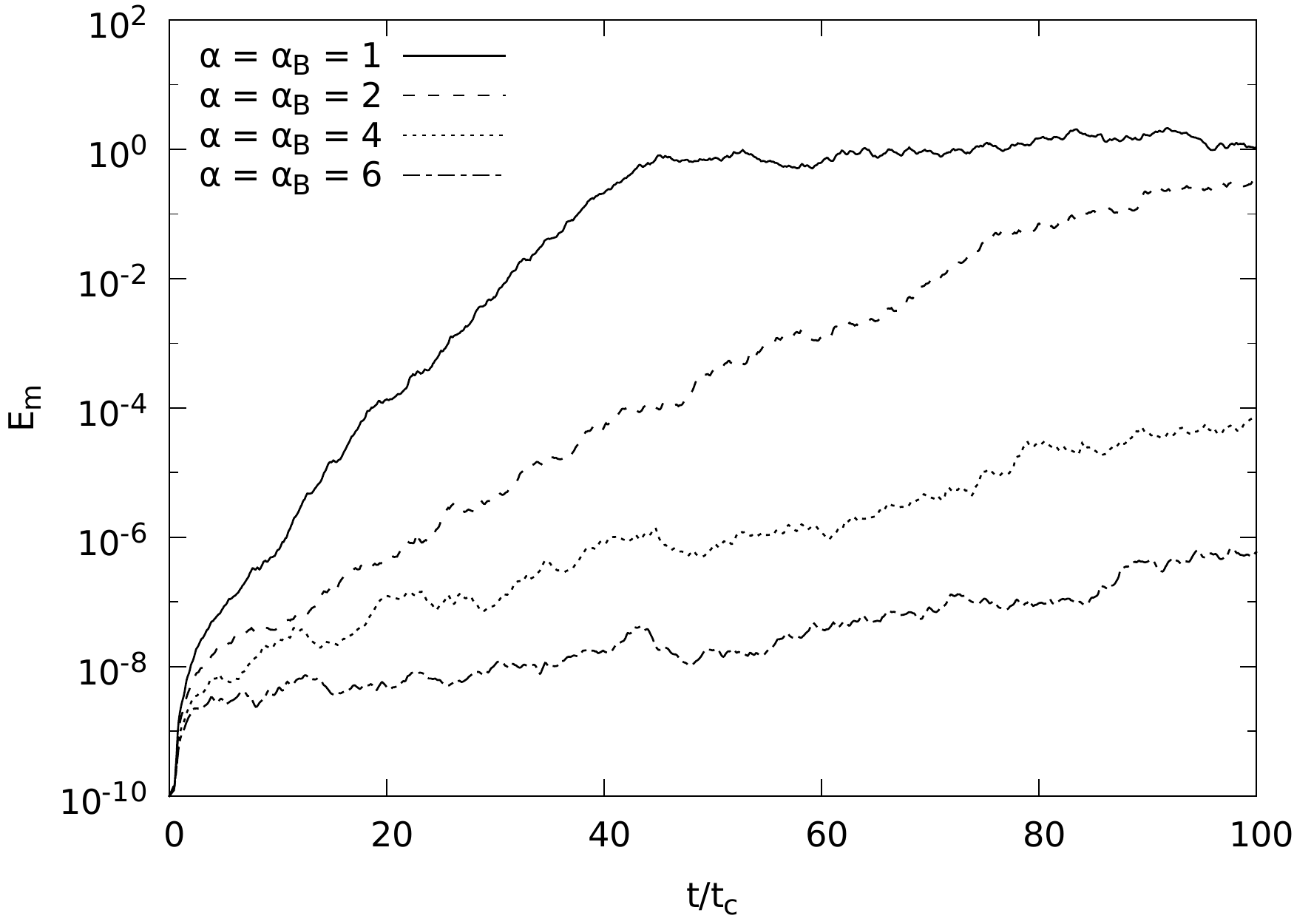} 
\caption{A $128^3$ {\sc Phantom} calculation where the artificial viscosity and resistivity parameters are systematically increased (no switches are used). With increasing dissipation, the growth rate decreases, producing the same behaviour consistent with changes in resolution.}
\label{fig:en_alphas}
\end{figure}

To further understand the dependence of the growth rate on resolution for {\sc Phantom}, a series of calculations were performed where the dimensionless parameters $\alpha$ and $\alpha_{\rm B}$ in the artificial viscosity and resistivity terms were fixed to different values. We found that the growth rate depended on the amount of artificial dissipation applied, and produced an effect equivalent to changing the resolution (Fig.~\ref{fig:en_alphas}). Since the dissipation in {\sc Phantom} is proportional to resolution, we conclude that the growth rates obtained in our comparison are consistent with the expected resolution scaling of the artificial dissipation terms. We comment that using $\alpha_{\rm B} = 8$ for a resolution of $128^3$ particles would produce a magnetic Reynolds number of ${\rm Rm} \sim 160$ (c.f. Appendix~\ref{sec:prandtl}), below estimates of the critical magnetic Reynolds number needed to support dynamo amplification \citep{schoberetal12a}.

\subsection{Magnetic energy saturation level}
\label{sec:satlevel}

The mean magnetic energy of the saturated magnetic field is approximately $2$--$4\%$ of the mean kinetic energy for all calculations (Table~\ref{tbl:energies}). This is consistent with theoretical predictions \citep{schoberetal15} and prior numerical studies \citep{federrathetal11, federrathetal14} of the small-scale dynamo in compressible, high Mach number turbulence. We note that incompressible turbulence has a higher saturation value, approaching $10$--$40$\% of the kinetic energy \citep{brandenburgetal96, hbd04, choetal09}.

The mean magnetic energy exhibits a trend of increasing with resolution, with the $512^3$ calculations twice as high as the corresponding calculations at $128^3$ (for both {\sc Flash} and {\sc Phantom}), though remains within the standard deviation. We note that the $512^3$ {\sc Phantom} calculation is averaged over a shorter time ($20t_{\rm c}$ compared to $50$--$70t_{\rm c}$), which is reflected by its smaller standard deviation. The $512^3$ {\sc Flash} calculation shows a long-term variation, with a $50\%$ increase in mean energy above $80t_{\rm c}$. This is reflected in the wider standard deviation in this calculation ($\sim1.0$ compared to $0.2$--$0.3$ in the $128^3$ and $256^3$ calculations). Overall, while the statistical ranges of mean energy overlap between resolutions, it appears that {\sc Phantom} yields higher mean magnetic energy during the saturation phase than {\sc Flash} at comparable resolution.

A set of {\sc Phantom} calculations were performed keeping the same artificial viscosity parameters but turning off the \citet{tp13} switch for artificial resistivity (i.e. using a constant artificial resistivity parameter, $\alpha_{\rm B} = 1$), thereby increasing the amount of resistive dissipation and lowering the magnetic Reynolds number.  This reduced the mean magnetic energy in the saturation phase at all three resolutions ($128^3$: 1.52 to 1.01, $256^3$: 2.31 to 1.32, $512^3$: 2.98 to 1.57). Considering that the Prandtl number in {\sc Flash} should be nearly constant with varying resolution (Appendix~\ref{sec:prandtl}), this suggests that the magnetic Reynolds number determines the saturation level of the magnetic field.

\subsection{Alfv\'enic Mach number}

\begin{figure}
 \centering
\includegraphics[width=\columnwidth]{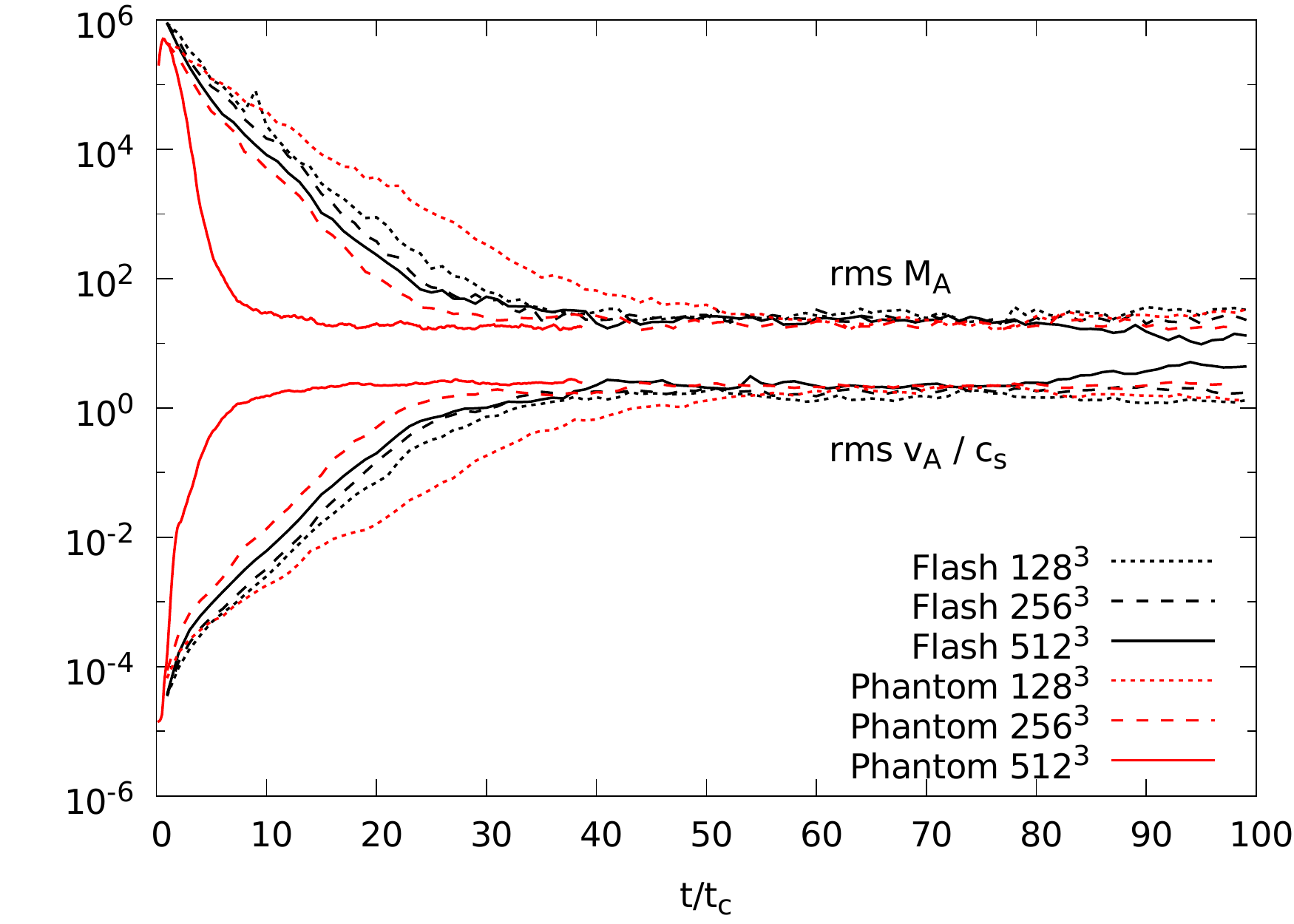}
\caption{Time evolution of the rms Alfv\'en speed and Alfv\'enic Mach number. For all calculations, the time averaged rms Alfv\'en speed in the saturation phase is $v_\text{A}\sim2 c_{\rm s}$. The rms Alfv\'enic Mach number is $M_\text{A}\sim20$, as calculated per cell or particle. This noticeably differs from the rms velocity divided by the rms Alfv\'en speed ($\sim5$).}
\label{fig:alfvenmach}
\end{figure}

Fig.~\ref{fig:alfvenmach} shows the time evolution of the rms Alfv\'en speed, $v_{\rm A}$, and rms Alfv\'enic Mach number, $\mathcal{M}_{\rm A}$ for all six calculations. The initial rms Alfv\'enic Mach number is $\mathcal{M}_{\rm A} \sim 10^6$, which decreases as the dynamo amplifies the magnetic energy and the rms Alfv\'en speed throughout the domain increases. In the saturation phase, the rms Alfv\'en speed is approximately twice the sound speed ($v_{\rm A} \sim 2 c_{\rm s}$) and the rms Alfv\'enic Mach number is $\mathcal{M}_{\rm A} \sim 20$. In other words, the turbulence remains super-Alfv\'enic even once the magnetic field has reached saturation. The Aflv\'enic Mach number in Fig.~\ref{fig:alfvenmach} is calculated by taking the rms of the local $\mathcal{M}_{\rm A}$ as calculated per grid cell or particle. This differs by a factor of four to that calculated by dividing the rms velocity ($10$) by the rms $v_{\rm A}$ ($2$), yielding $\sim5$, suggesting a correlation between the velocity and magnetic field.

\subsection{Magnetic energy power spectra}
\label{sec:magspectra}

\begin{figure}
 \centering
\includegraphics[width=\columnwidth]{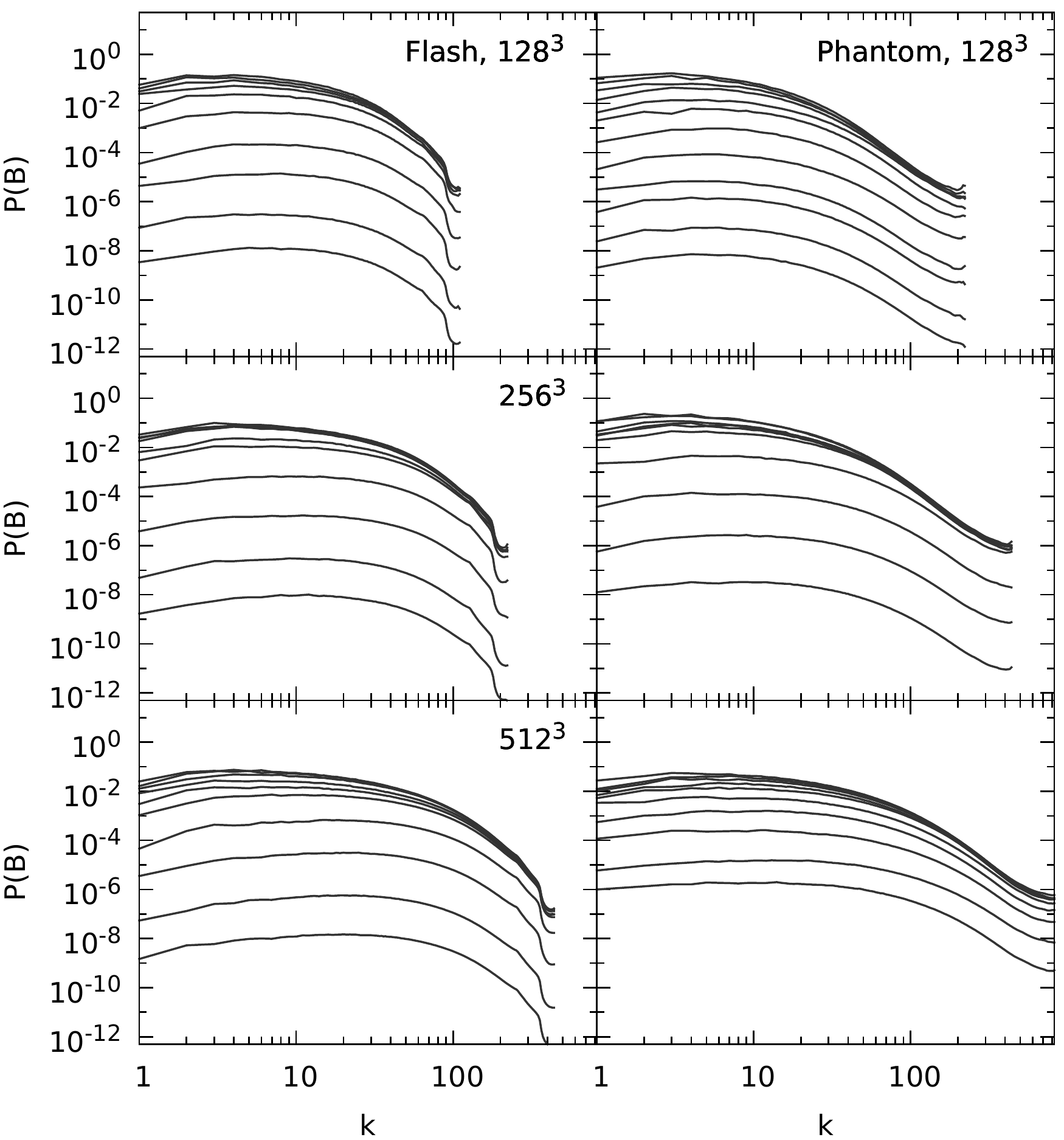}
\caption{Spectra of the magnetic energy during the growth phase for {\sc Flash} (left) and {\sc Phantom} (right) for resolutions of $128^3$, $256^3$, and $512^3$ (top to bottom). Each spectral line is sampled at intervals of $5t/t_{\rm c}$ up to $t/t_{\rm c}=50$, except for the $512^3$ {\sc Phantom} run which is sampled every $t/t_{\rm c}$ (from $t/t_{\rm c}=2$--$12$). The magnetic field grows equally at all spatial scales for all calculations. The saturation of the magnetic energy occurs first at the smallest scales, with a time delay before the larger scales saturate.}
\label{fig:growthspectra}
\end{figure}

That the total magnetic field is growing in strength --- and not just in isolated regions --- may be quantified by examining the power spectra of the magnetic energy, ${\rm P}(B)$. The magnetic energy spectra during the growth phase for the six calculations is presented in Fig.~\ref{fig:growthspectra}. The magnetic energy can be seen to grow uniformly at all spatial scales in all six calculations (indicated by the translation of the power spectrum along the y-axis in the plots with minimal change in the shape), behaviour consistent with the small-scale dynamo \citep{mcm04, schekochihinetal04c, bs05, choetal09, bp13, federrathetal14}. 

All of the spectra have the same general shape, with a decrease in spectral energy at and above the driving scale ($k\le3$) and a more-or-less flat spectrum (${\rm P}(B) \approx$ constant) between $3<k<10$ for the $128^3$ calculations, extending to $k\sim20$ and $k\sim40$ for the $256^3$ and $512^3$ calculations. The dissipation range in the {\sc Phantom} results extends further to smaller scales than the {\sc Flash} results for a particular resolution. The maximum in the magnetic energy spectrum in both codes occurs at high wavenumbers, as expected for small-scale dynamos \citep{cv00,bss12}, occurring around the high $k$ end of the `relatively' flat region of the spectra.

Fig.~\ref{fig:growthspectra} shows that the magnetic energy saturates first at small scales. This is characteristic of the small-scale dynamo since this is where magnetic energy is being injected \citep{choetal09}. It is expected that the magnetic energy will grow linearly at this stage, though for Burgers turbulence, which is closer to the regime our simulations are in, it is expected that the magnetic energy growth will be closer to quadratic \citep{schleicheretal13}.  This slow growth phase lasts until the reverse cascade of magnetic energy saturates all spatial scales. This turnover in magnetic energy growth may be clearly seen in the $128^3$ and $512^{3}$ {\sc Phantom} growth curves in Fig.~\ref{fig:en_mag}. 

\begin{figure}
 \centering
\includegraphics[width=\columnwidth]{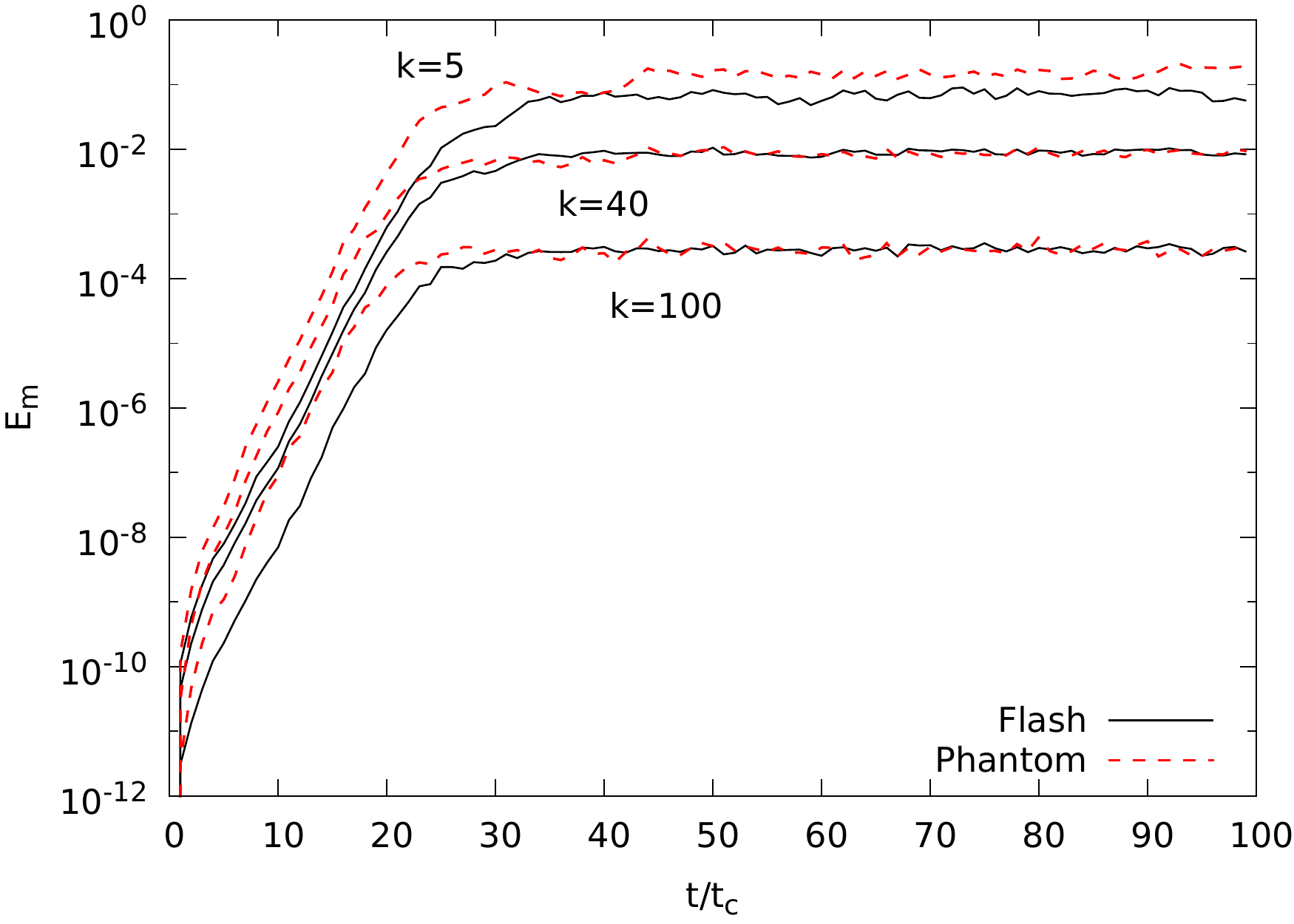}
\caption{Spectra of the magnetic energy at the $k=5$, $40$, and $100$ bands as a function of time for the $256^3$ resolution calculations of {\sc Flash} (black lines) and {\sc Phantom} (red dashed lines). The growth rate at these different wave numbers is nearly identical. The saturation level is the same between the two codes for $k=40$,$100$, with {\sc Phantom} containing $\sim2$ times as much energy in the large-scale $k=5$ band.}
\label{fig:kbands}
\end{figure}

Fig.~\ref{fig:kbands} shows a cross section of the power spectrum evolution at $k=5$, $40$, and $100$ for the $256^3$ calculations. These scales were chosen to represent large, medium, and small-scale structure. This shows that the magnetic field grows in the same manner at all scales in both codes. It is also evident that the magnetic field enters the slow growth phase first at high wavenumbers.

\begin{figure}
 \centering
\includegraphics[width=0.99\columnwidth]{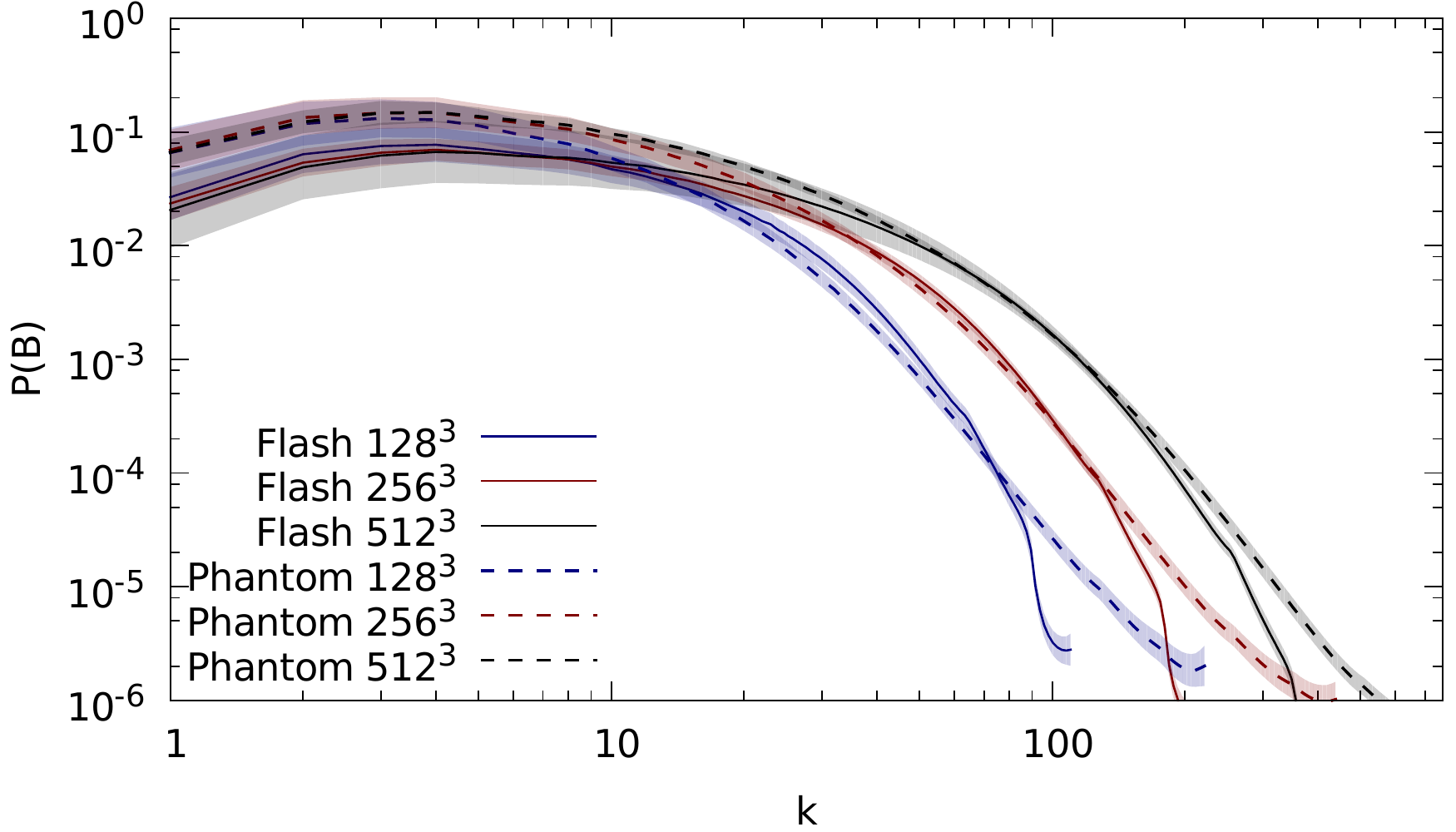}
\caption{Time averaged spectra of the magnetic energy in the saturation phase for {\sc Flash} (solid lines) and {\sc Phantom} (dashed lines) at resolutions of $128^3$ (blue), $256^3$ (red), and $512^3$ (black). Shaded regions represent the standard deviation. The {\sc Phantom} calculations systematically contain more magnetic energy (approximately $2\times$) in large-scale structure ($k<10$) compared to {\sc Flash}, and have an extended tail at high $k$ due to the adaptive resolution.}
\label{fig:satspect}
\end{figure}

Fig.~\ref{fig:satspect} shows the time-averaged spectra of the magnetic energy from all six calculations in the saturation phase, with the shaded regions showing one standard deviation of the time-average. In each case 50 spectra have been averaged over a minimum of $50 t_{\rm c}$, with the exception of the $512^3$ {\sc Phantom} calculation which has been averaged over $20 t_{\rm c}$. The spectra of {\sc Flash} and {\sc Phantom} are similar in shape, except that the {\sc Phantom} calculations contain approximately twice as much magnetic energy in large-scale structure ($k<10$). This is consistent with the higher mean magnetic energy in the {\sc Phantom} calculations in Table~\ref{tbl:energies}, indicating that this energy is stored in the largest scales of the field.

The peak of the magnetic energy spectra for both codes is at $k\sim3$--$4$, occurring just above the driving scale. As the resolution is increased, both codes extend the spectra further toward small scales. The {\sc Flash} power spectra drop sharply at the Nyquist frequency, while the {\sc Phantom} power spectra reach higher wavenumbers than {\sc Flash} for the same number of resolution elements. While, in SPH, the smoothing kernel will distribute power to higher wavenumbers, the {\sc Phantom} calculations have adaptive resolution that reach $4$--$8\times$ that of the {\sc Flash} calculation in the densest regions. For that reason, the {\sc Phantom} power spectra have been analysed on a grid that is twice the resolution of the {\sc Flash} grid (see Appendix~\ref{sec:gridinterp} for why this grid resolution was chosen), and it is expected that these power spectra correspond to resolved structures.

\begin{figure}
 \centering
\includegraphics[width=0.99\columnwidth]{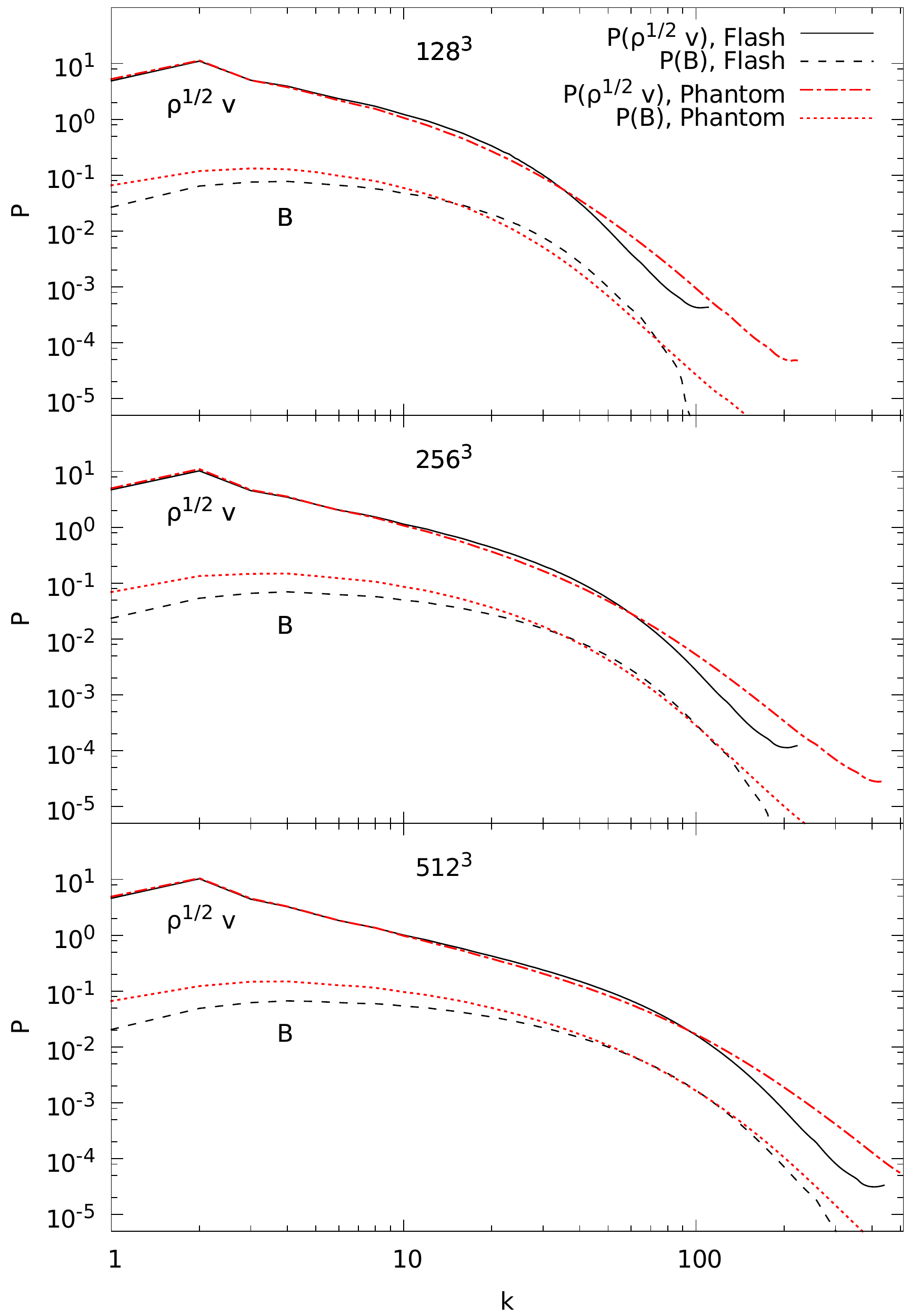}
\caption{Time averaged kinetic ($\rho^{1/2} v$) and magnetic ($B$) energy spectra in the saturated phase for {\sc Flash} (black lines) and {\sc Phantom} (red lines). As the resolution is increased, the kinetic and magnetic energy spectra extend to higher wavenumbers, with the magnetic energy lower than kinetic energy at all wavenumbers.}
\label{fig:spectmagkin}
\end{figure}

Fig.~\ref{fig:spectmagkin} compares the magnetic spectra to the kinetic energy spectra. It is characteristic for the small-scale dynamo for the peak in the magnetic energy spectrum to be at a wavenumber just above the peak in the kinetic energy spectrum \citep{cv00, bss12}. This is clearly seen in Fig.~\ref{fig:spectmagkin}. The sharp peak at $k=2$ in the kinetic energy spectra is due to the driving force, and for all resolutions the peak of the magnetic energy spectra occurs just above this scale ($k\sim3$--$4$).

The magnetic energy is lower than the kinetic energy at all wave numbers. \citet{brandenburgetal96, schekochihinetal04c, mcm04, hbd04, choetal09} found that the magnetic energy spectrum overtook the velocity spectrum, ${\rm P}(v)$, at high $k$ in incompressible and subsonic calculations. We similarly found that the magnetic energy spectra of our calculations exceeded the velocity spectra at high $k$ (without taking into account the density field), but since we are dealing with compressible, supersonic turbulence, we instead investigate the kinetic energy spectrum, i.e., ${\rm P}(\rho^{1/2} v)$. In this case, the magnetic energy spectrum is below the kinetic energy spectrum at all wavenumbers for both {\sc Flash} and {\sc Phantom}.

\subsection{PDFs of $B^{2}$}
\label{sec:pdfsB2}

 \begin{figure*}
\centering
\includegraphics[width=0.99\textwidth]{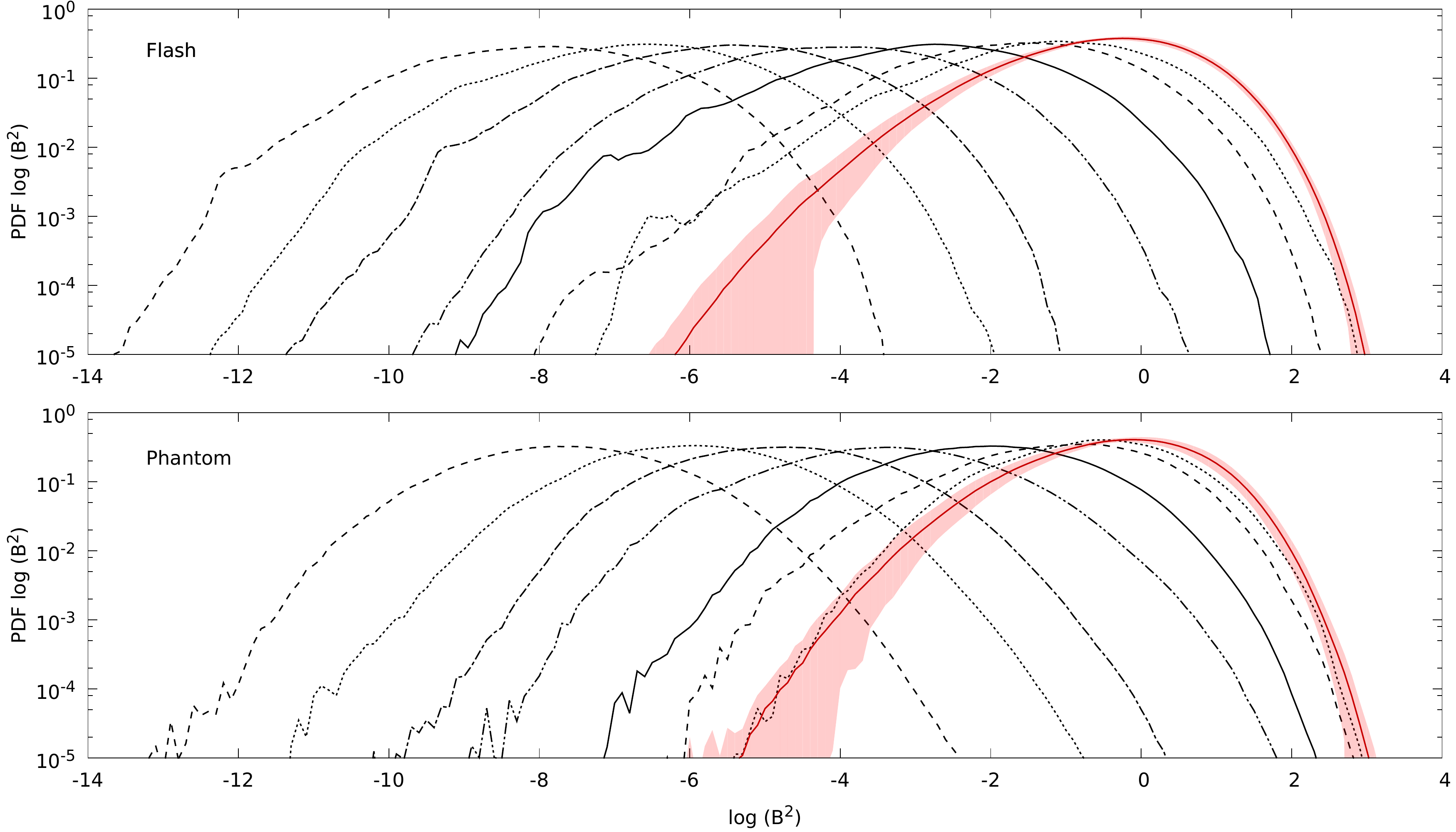}
\caption{PDF of $\log(B^2)$ during the growth phase, with the red line time averaged during the saturation phase. The top panel shows the {\sc Flash} calculation, with the bottom panel the {\sc Phantom} calculation. The PDF has a log-normal distribution during the growth phase, maintaining its width while the peak smoothly translates to higher magnetic field strengths. As the strongest magnetic fields saturate, the PDFs become lop-sided. Both PDFs in the saturation phase have similar peaks and high-end tails, with {\sc Flash} exhibiting a slightly extended low-end tail.}
\label{fig:bsqpdf}
\end{figure*}

\begin{figure}
\centering
\includegraphics[width=0.99\columnwidth]{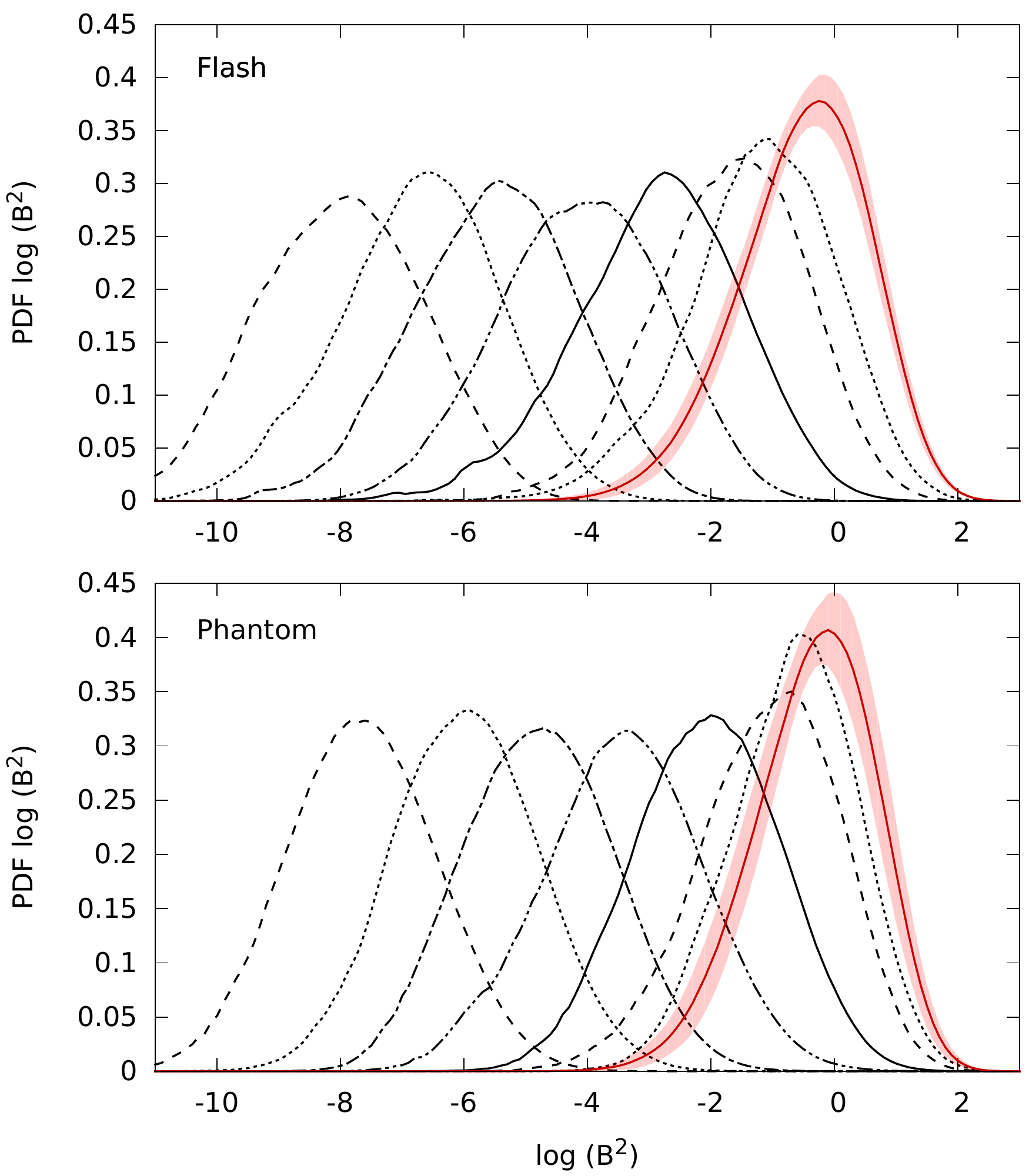}
\caption{PDF of $\log(B^2)$ during the growth phase, with the red line time averaged during the saturation phase. This is equivalent to Fig.~\ref{fig:bsqpdf} but on a linearly scaled plot. In the saturation phase, the distribution is skewed with smaller deviation of magnetic field strengths.}
\label{fig:bsqpdf-linear}
\end{figure}

Fig.~\ref{fig:bsqpdf} shows the time evolution of the probability distribution function (PDF) of $B^2$ for the $256^3$ calculations. The instantaneous PDFs are shown from $t/t_{\rm c}=4$--$28$ at intervals of $\Delta t = 4 t_{\rm c}$, with the time-averaged PDF during the saturation phase given by the red line with the shaded region representing the standard deviation of the time-averaging. The shape of the PDFs remain mostly log-normal during the growth phase. As the dynamo amplifies the magnetic field, the PDFs maintain their width and shape, with the peak simply translating to higher magnetic field strengths. In other words, the distribution of magnetic fields does not broaden over time, but all magnetic field strengths are increased uniformly such that the PDFs maintain their shape and width while the mean increases \citep{schekochihinetal04b, schekochihinetal04c}.

The PDFs become distorted during the slow growth phase as the magnetic field approaches saturation. Once the dynamo enters the slow growth regime, it is no longer able to amplify the strongest magnetic fields due to the back-reaction of the Lorentz force. Thus, the high-end tail of the distribution anchors in place, leading to a `squeezed' distribution as the magnetic fields in the peak and low-end tail continue to increase \citep{schekochihinetal04b, schekochihinetal04c}. This produces a lop-sided distribution and both codes show this behaviour as the magnetic field saturates. This may also be seen in Fig.~\ref{fig:bsqpdf-linear} which shows the PDF of $B^2$ on a linear scale. 

In the saturation phase, the distributions peak at similar magnetic field strengths, agreeing to within 10\% on the maximum of the peak. They have similar ranges on the high-end tail of the distribution, and the maximum magnetic field achievable agrees to within 10\%. The low-end tail extends further for {\sc Flash}, and has a larger standard deviation. From the linearly scaled plots of Fig.~\ref{fig:bsqpdf-linear}, it is seen that the probability of being at the mean magnetic field strength increases by $\sim20\%$ once the magnetic field has saturated, corresponding to a reduced variance in the distribution of magnetic field strengths. This occurs due to the saturation of the strongest magnetic fields \citep{schekochihinetal04b, schekochihinetal04c}. 

Fig.~\ref{fig:bsqpdf} additionally shows that {\sc Flash} is able to sample lower magnetic field strengths compared to \textsc{Phantom}, which was similarly noted by \citetalias{pf10} in the density PDFs and was attributed to the better weighting of resolution elements towards low density regions in the grid code. The extended low-end tail of the PDFs of $B^2$ for the grid code is consistent with this, since the low density regions are expected to contain weaker magnetic fields.

\subsection{Density PDFs}
\label{sec:densitypdfs}

\vspace{3mm}

For supersonic turbulence, the PDF of $s \equiv \ln (\rho / \rho_0)$ follows a log-normal distribution \citep[e.g.,][]{vs94, pnj97, pvs98, np99, klessen00, ls08, fks08, federrathetal10, pf10, knw11, fk13}. For the motivation behind this choice of variable, $s$, see \citet{vs94} and \citet{fks08}. This log-normal distribution is a consequence of the density at a location being perturbed randomly and independently over time, which according to the central limit theorem, will converge to a log-normal distribution \citep{papoulis84, vs94}. Other processes may affect the shape of the PDF. Higher Mach numbers broaden the width of the distribution \citep{klb07, ls08, federrathetal10, pfb11, konstandinetal12}, self-gravity has been demonstrated to add power-law tails at high densities \citep{klessen00, lkmm03, knw11, fk13, girichidisetal14}, non-isothermal equations of state can introduce power-law tails at high and low densities \citep{pvs98, lkmm03, fb15, nfs15}, different forcing mechanisms (compressive vs solenoidal) influence the shape of the distribution \citep{fks08, federrathetal10, federrath13}, and, relevant for the current discussion, magnetic fields can narrow the width of the distribution by effectively decreasing the compressibility of the gas  \citep{pn11, collinsetal12, molinaetal12, fk13}.

\citetalias{pf10} compared the PDF of $s$ for hydrodynamic turbulence, finding that SPH yielded a log-normal distribution as expected. There were differences when compared to PDFs obtained with grid-based methods. In particular, the density PDF obtained with SPH extended further to higher densities, but conversely had a shorter low-end tail. This was a consequence of the resolution in SPH being adaptive in density, that is, high density regions have more resolution. Now we consider how the addition of magnetic fields may affect the density PDF in SPMHD.

\begin{figure}
 \centering
\includegraphics[width=\columnwidth]{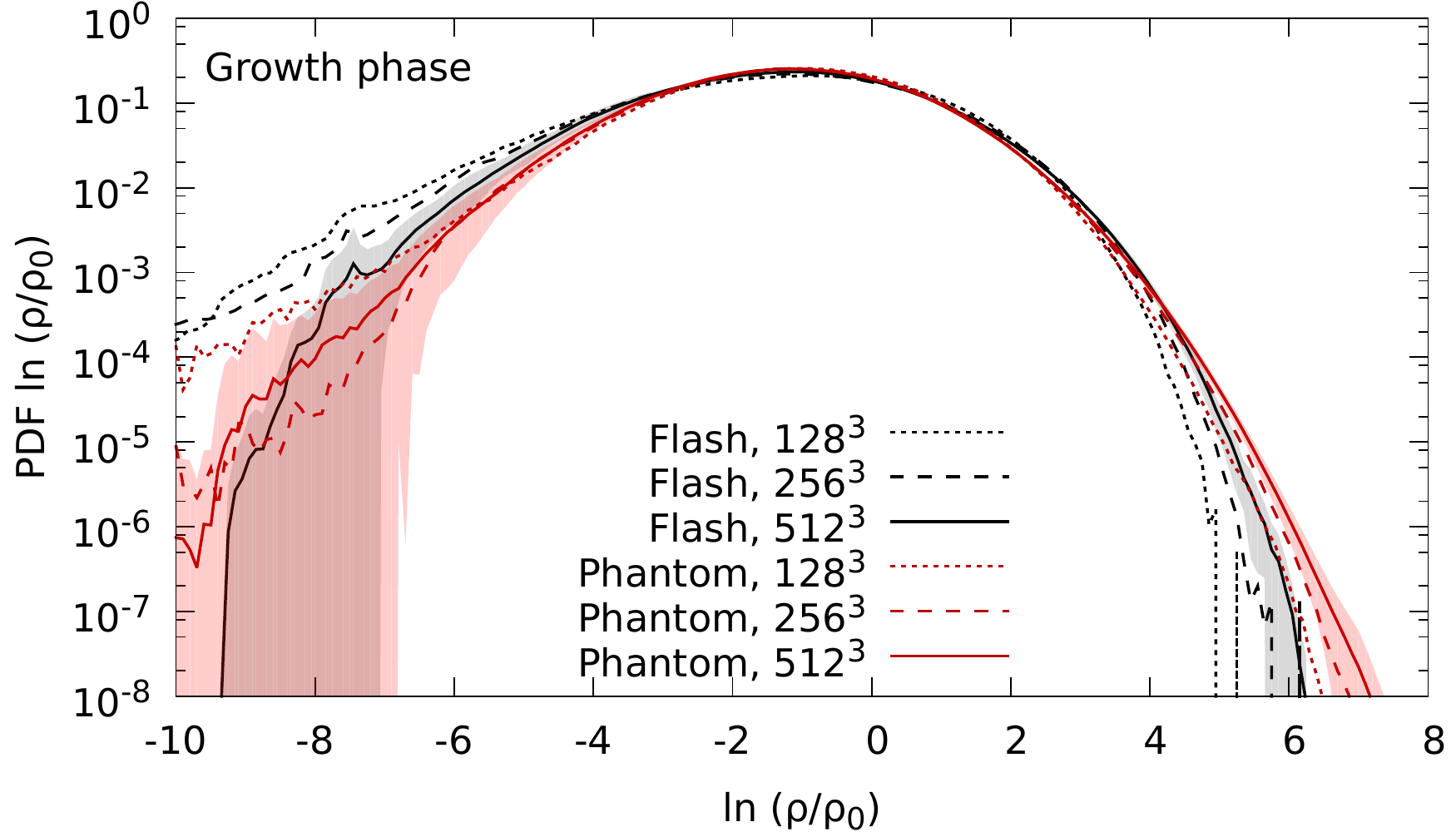} \\
\includegraphics[width=\columnwidth]{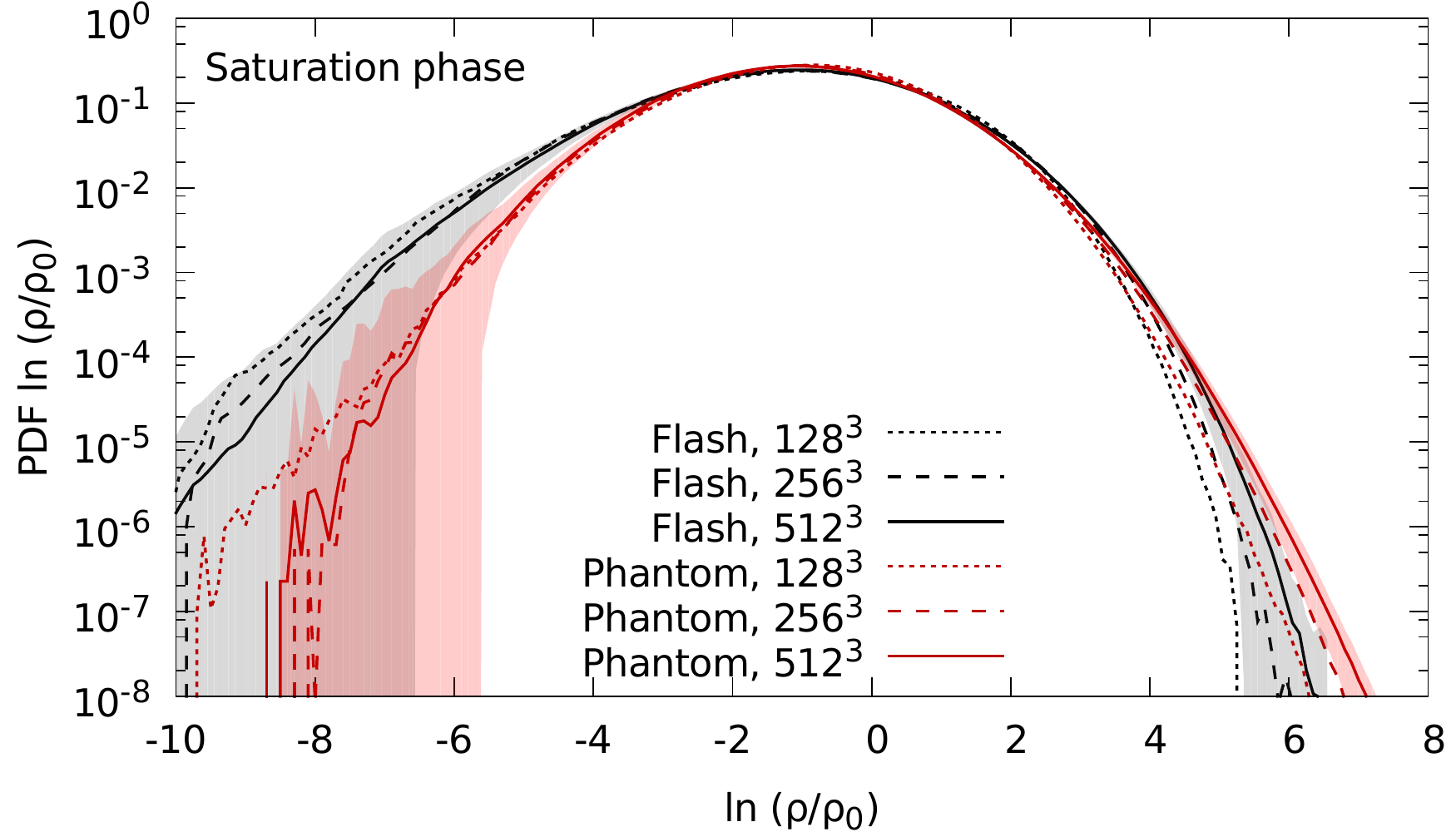}
\caption{Time averaged density PDFs during the growth phase (top panel, for $t/t_{\rm c}=2$--$10$) and during the saturation phase (bottom panel, for $t/t_{\rm c}=30$--$100$, only $t\ge50$ for the $128^3$ {\sc Phantom} calculation).  The peaks and high end tail of the PDF are similar for both cases, but the low density tail is less extended when the magnetic field has reached saturation.}
\label{fig:densitypdfs}
\end{figure}

\begin{figure}
 \centering
\includegraphics[width=\columnwidth]{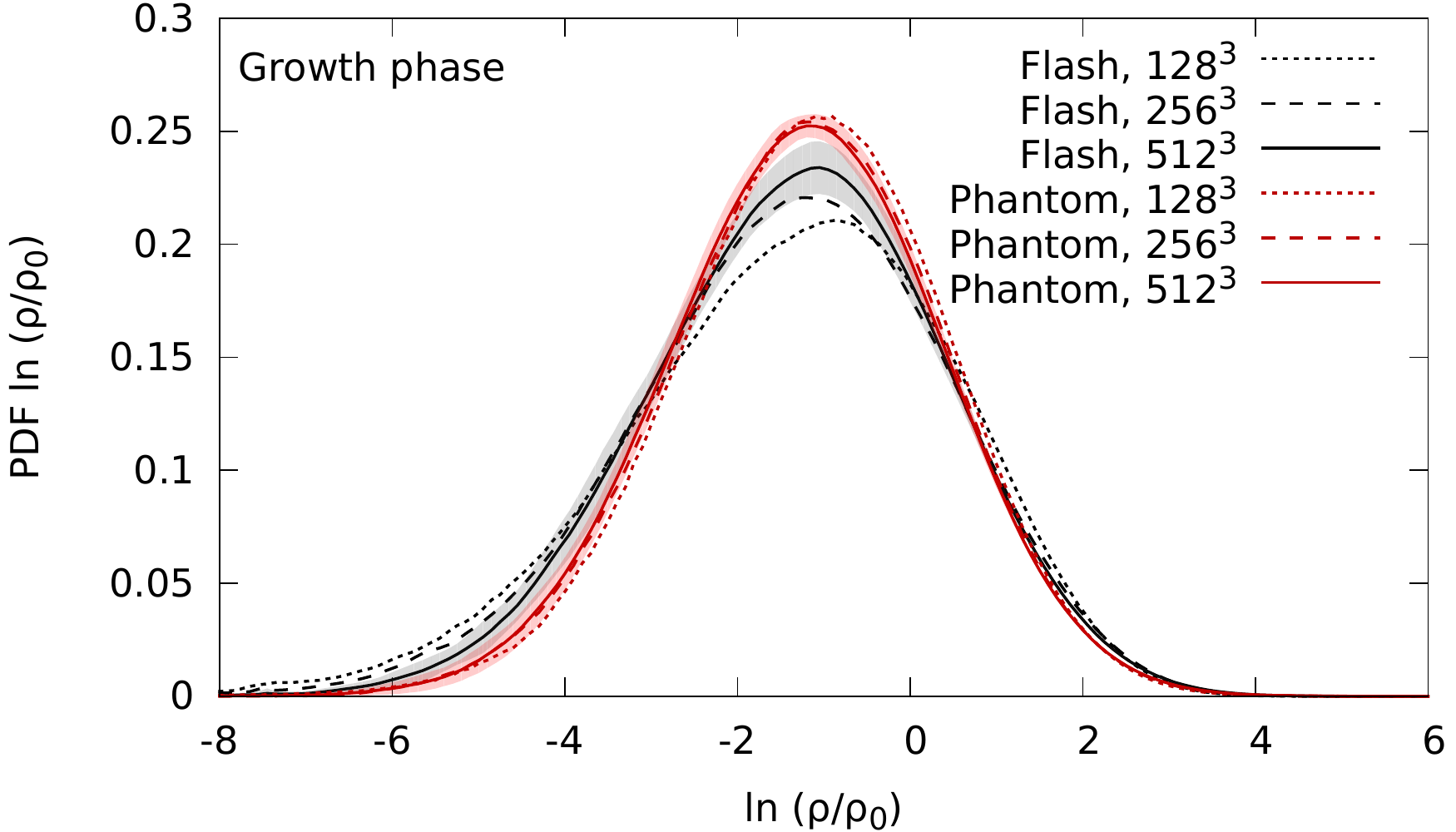} \\
\includegraphics[width=\columnwidth]{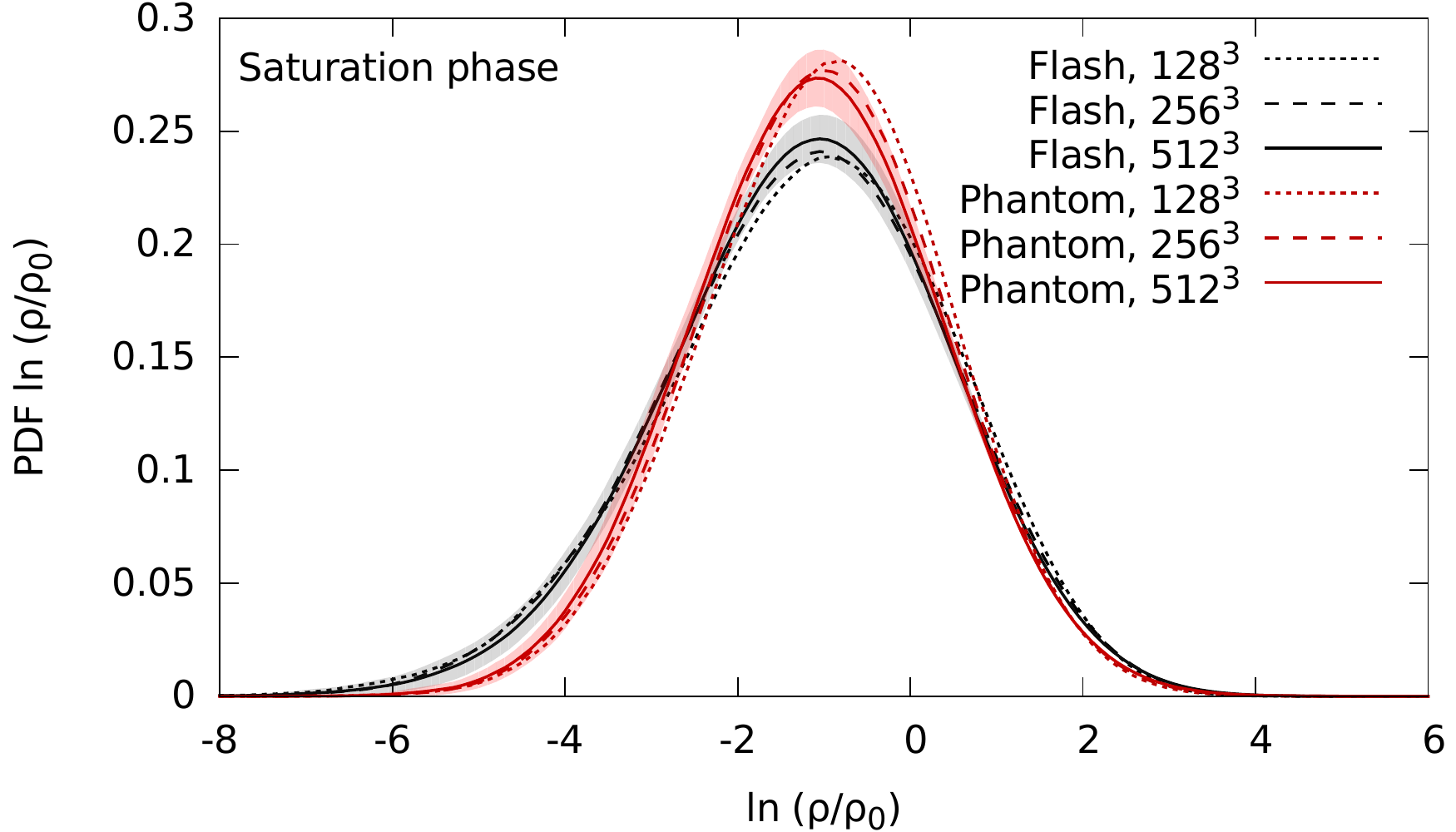}
\caption{Time averaged density PDFs during the growth phase (top panel, for $t/t_{\rm c}=2$--$20$) and during the saturation phase (bottom panel, for $t/t_{\rm c}=30$--$100$, only $t/t_{\rm c}\ge50$ for the $128^3$ {\sc Phantom} calculation).  This is equivalent to Fig.~\ref{fig:densitypdfs} but on a linearly scaled plot.  The peaks and high end tail of the PDF are similar for both cases, but the low density tail is less extended when the magnetic field has reached saturation.}
\label{fig:densitypdfs-linear}
\end{figure}

Fig.~\ref{fig:densitypdfs} compares the PDFs of the density contrast, $s$, during the growth phase, while the magnetic field is dynamically weak, to the saturation phase when the magnetic field is at its strongest. The PDFs in the growth phase were time-averaged during the first half of the growth phase while $E_{\rm m} < 10^{-4}$ (excluding the initial transient growth). This allows for statistical averaging over a number of crossing times while the magnetic field is still dynamically weak. The PDFs in the saturation phase were averaged over at least $50 t_{\rm c}$, except the $512^3$ {\sc Phantom} calculation averaged over $20t_{\rm c}$. The standard deviation from the time averaging is shown for the highest resolution calculations by the shaded regions (black for {\sc Flash}, red for {\sc Phantom}).
 
Both codes show PDFs close to a log-normal distribution in both the growth and saturation phases. {\sc Flash} can be seen to sample a lower range of densities, while {\sc Phantom} samples a higher range. This behaviour is similar to that found by \citetalias{pf10}, and occurs because {\sc Phantom} uses adaptive resolution based on the density. In the saturation phase, the extra support from magnetic pressure reduces the low-end tail of the distribution, making it more log-normally distributed, consistent with previous findings \citep{klb07, ls08, pfb11, molinaetal12, fk13}. The peak and high-end tail of the distribution remain quite similar during both the growth and saturation phases. Fig.~\ref{fig:densitypdfs-linear} shows the PDFs of $s$ on a linear scale. As before, the extended low-end tail for {\sc Flash} is visible, but from this it is also clear that the mean of the distribution is higher for {\sc Phantom}, meaning that the distribution is slightly narrower. This was also noted by \citetalias{pf10}.

\section{Conclusion}
\label{sec:turbsummary}

We have performed a comparison of particle-based SPMHD methods using the {\sc Phantom} code with results from the grid-based code {\sc Flash} on the small-scale dynamo amplification of magnetic fields. The calculations used supersonic turbulence driven at rms velocity Mach 10 in an isothermal fluid contained in a periodic box. The initial magnetic field was uniform and had an energy $12$ orders of magnitude smaller than the mean kinetic energy of the turbulence. The small-scale dynamo amplification of the magnetic field was followed for $10$ orders of magnitude in energy. 

The three phases of small-scale dynamo amplification were modeled: exponential growth phase, slow linear or quadratic growth phase as the magnetic field started to saturate, and the fully saturated phase of the magnetic field. We considered the exponential growth rate of magnetic energy, saturation level of the magnetic energy, the power spectrum of magnetic, and the PDFs of the magnetic and density fields. 

Our main conclusion is that SPMHD can successfully reproduce the exponential growth and saturation of an initially weak magnetic field via the small-scale dynamo in magnetised, supersonic turbulence. Our simulation results are summarised as follows:

\begin{itemize}
\item Both methods exhibited similar qualitative behaviour. The initially weak magnetic field was exponentially amplified at a steady rate over a period of tens of turbulent crossing times, with a slow turnover in magnetic energy corresponding to the slow growth phase. The regions of strongest magnetic field correlated with the high density regions in the results obtained from both methods. The magnetic energy was amplified by 10 orders of magnitude, saturating when it was $2$--$4$\% of the kinetic energy. 

\item The growth rate of magnetic energy in the {\sc Flash} calculations varied only slightly with resolution (5--10\%), whereas the {\sc Phantom} calculations nearly doubled the growth rate with each factor of two increase in resolution. This was found to be consistent with the resolution scaling of the artificial viscosity and artificial resistivity used to capture shocks and discontinuities in the magnetic field, as these set the level of numerical dissipation, and subsequently, the growth rate.

\item {\sc Phantom} is more computationally expensive than {\sc Flash}, with the {\sc Phantom} calculations taking approximately $30\times$ more cpu-hours than the {\sc Flash} calculations at comparable resolution. The $256^3$ {\sc Phantom} calculation required as many cpu-hours as the $512^3$ {\sc Flash} calculation, consistent with the purely hydrodynamic results of \citetalias{pf10}, meaning that adding MHD to SPH adds negligible computational expense. 

\item In both sets of calculations, the magnetic energy spectrum grew uniformly at all spatial scales during the exponential growth phase. The spectra saturated first at the smallest scale, corresponding to the scale at which energy is injected into the magnetic field, after which there was a phase of slow growth as the magnetic energy spectra slowly saturated at larger scales. This behaviour is consistent with the small-scale dynamo. The magnetic energy spectra in the saturation phase are relatively flat on large scales, peaking around $k\sim3$--$5$. The magnetic energy spectra of {\sc Phantom} in the saturation phase contain twice as much magnetic energy at large scales compared to the spectra from {\sc Flash}, reflected by the higher mean magnetic energy.

\item The distribution of magnetic field strengths had similar behaviour in both sets of calculations, both during the exponential growth and saturation phases. During the growth phase, both codes produced a log-normal PDF of $B^2$ which maintained its width and shape over time, but with the peak increasing to higher field strengths. As the magnetic field approached saturation, the PDF of $B^2$ deviated from a log-normal distribution. The high-end tail remained fixed during the slow growth phase, but the peak and low-end tail continued increasing leading to a lop-sided distribution. Both the {\sc Phantom} and {\sc Flash} results evidenced this behaviour, and agreed on the maximum magnetic field strength and most probable magnetic field strength to within 10\% each. {\sc Flash} had an extended low-end tail, resulting from SPMHD being adaptive in resolution with density.

\item The density PDF was examined during the growth phase before the magnetic field became dynamically important, and during the saturation phase when the magnetic field was strongest. Both codes yielded density PDFs which were log-normal, with possible suggestions that the low-end tail of the distribution was reduced in the saturation phase. The density PDFs of {\sc Phantom} extended further to high densities, whereas the density PDFs of {\sc Flash} extended further to low densities. This is a consequence of the density adaptive resolution of SPH.
\end{itemize}

\section*{Acknowledgments}

We thank the anonymous referee for their comments which have improved the quality of this paper. TST is supported by a CITA Postdoctoral Research Fellowship, and was supported by Endeavour IPRS and APA post-graduate research scholarships during which this work was completed. DJP is supported by ARC Future Fellowship FTI130100034, and acknowledges funding provided by the ARC's Discovery Projects (grant no.~DP130102078). CF acknowledges funding provided by the Australian Research Council's Discovery Projects (grants~DP130102078 and~DP150104329). This research was undertaken with the assistance of resources provided at the Multi-modal Australian ScienceS Imaging and Visualisation Environment (MASSIVE) through the National Computational Merit Allocation Scheme supported by the Australian Government. We gratefully acknowledge the J\"ulich Supercomputing Centre (grant hhd20), the Leibniz Rechenzentrum and the Gauss Centre for Supercomputing (grants pr32lo, pr48pi and GCS Large-scale project 10391), the Partnership for Advanced Computing in Europe (PRACE grant pr89mu), and the Australian National Computing Infrastructure (grant ek9). Fig.~2 and 3 were created using the interactive SPH visualisation tool {\sc Splash} \citep{splash}.

\bibliographystyle{mnras}
\bibliography{fullbib}

\appendix

\section{Interpolating particle data to a grid}
\label{sec:gridinterp}

The power spectra of kinetic and magnetic energy of the {\sc Phantom} calculations are computed by interpolating the particle data to a grid using an SPH kernel weighted summation over neighbouring particles. We have investigated whether using mass weighted or volume weighted interpolation changes the results. Furthermore, we have tested grids of varying resolution to find the optimal grid resolution to properly represent the magnetic field.

\begin{figure}
\centering
 \includegraphics[width=\columnwidth]{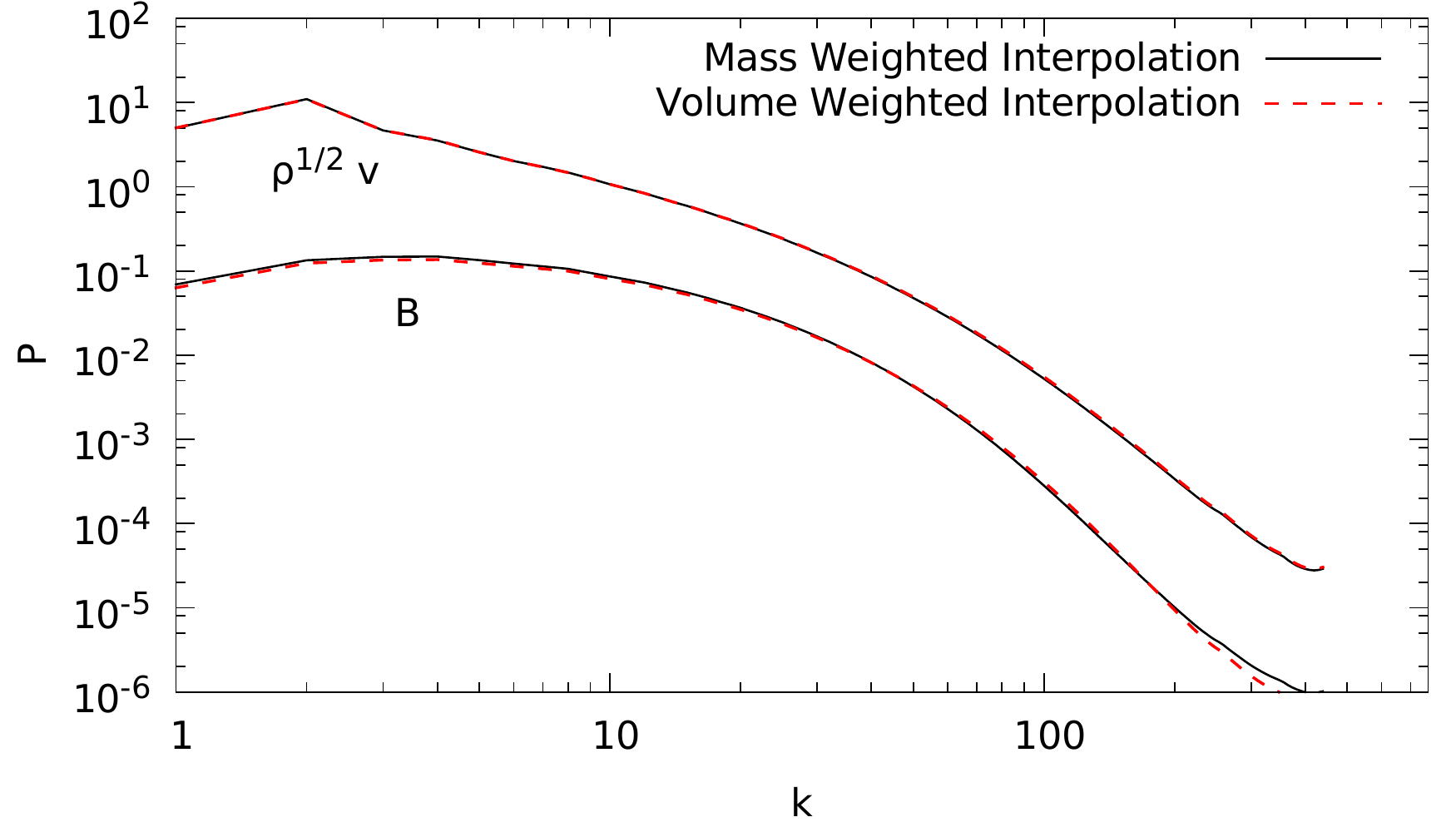}
\caption{Time averaged kinetic ($\rho^{1/2} v$) and magnetic ($B$) energy spectra of the $256^3$ particle {\sc Phantom} calculation interpolated to a $512^3$ grid using mass weighted and volume weighted interpolation. Both approaches yield the same result.}
\label{fig:interpspect}
\end{figure}

\begin{figure}
\centering
 \includegraphics[width=\columnwidth]{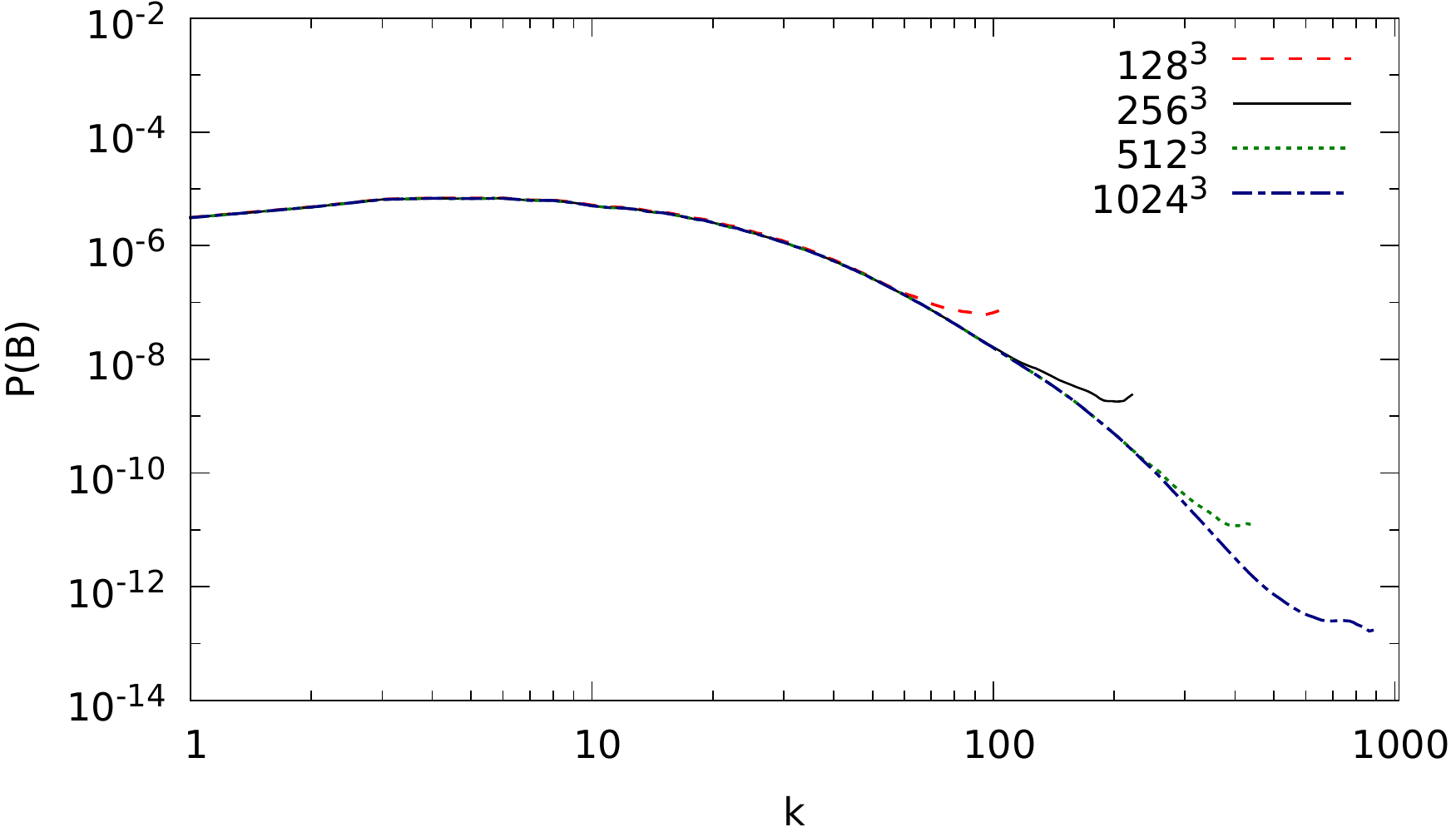}
\caption{Conversion of a $128^3$ particle {\sc Phantom} snapshot to grids of resolutions from $128^3$ to $1024^3$ grid points. Each resolution agrees well on the large-scale structure, but captures more of the small-scale structure as the resolution is increased. We find that the $256^3$ resolution grid sufficiently captures the total magnetic energy, therefore we choose grids with double the number of grid points as particles for our analysis.}
\label{fig:gridreso}
\end{figure}

The volume weighted interpolation of a quantity $A$ (in this case, the magnetic field $\bm{B}$) may be computed according to
\begin{equation}
 \bm{B}(\bm{r}) = \frac{\sum_b \dfrac{m_b}{\rho_b} \bm{B}_b W(\vert \bm{r} - \bm{r}_b\vert, h_b)}{\sum_c \dfrac{m_c}{\rho_c} W(\vert \bm{r} - \bm{r}_c \vert, h_c)} .
\end{equation}
The denominator is the normalization condition. A mass weighted interpolation may be computed as
\begin{equation}
 \bm{B}(\bm{r}) = \frac{\sum_b m_b \bm{B}_b W(\vert \bm{r} - \bm{r}_b \vert, h_b)}{\sum_c m_c W(\vert \bm{r} - \bm{r}_c \vert, h_c)} .
\end{equation}

Fig.~\ref{fig:interpspect} shows the kinetic and magnetic energy spectra for the $256^3$ {\sc Phantom} calculation computed from a $512^3$ grid using volume weighted and mass weighted interpolations, respectively. The spectra between the two interpolation methods are nearly indistinguishable, differing from each other by less than 1\% at all $k$ and deviating only near the resolution scale. We conclude that either approach is acceptable, and in this work we have used the mass weighted interpolation.

The smoothing length in our calculations can decrease by up to $8\times$ in the highest density regions, therefore we have tested the effect of different grid resolutions on the magnetic spectra. Fig.~\ref{fig:gridreso} shows magnetic energy spectra from a $128^3$ particle {\sc Phantom} calculation interpolated to grids with resolutions of $128^3$ to $1024^3$. Our results show that the large-scale structure ($k<50$) is nearly identical at all grid resolutions, with the spectra differing on the order of $0.1\%$ at each $k$-band. The only difference is that the spectra extend to higher $k$ as the resolution is increased. We find that the magnetic energy contained on the $128^3$ grid differs by $1\%$ of the energy contained on the particles, while the $256^3$ grid resolution differs by only $0.1\%$. Higher resolutions only minimally change the energy content of the magnetic field. Our conclusion is that a grid with double the resolution of the {\sc Phantom} calculation is sufficient for computing the magnetic energy spectrum accurately.

\section{Effective magnetic Prandtl numbers in grid and particle methods}
\label{sec:prandtl}

\begin{figure*}
\centering
\includegraphics[width=0.75\textwidth]{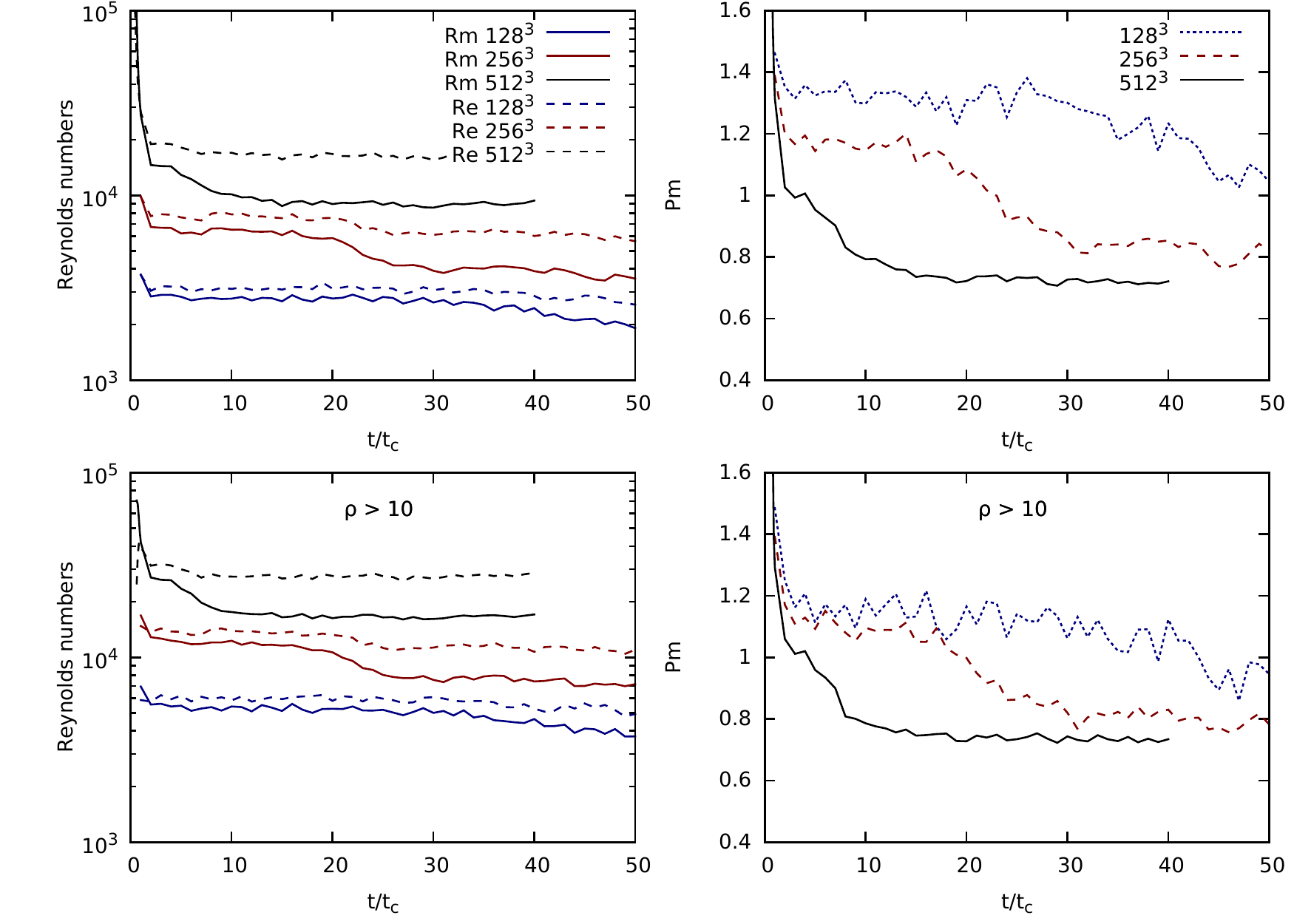}
\caption{The kinetic and magnetic Reynolds numbers (left plots) and Prandtl numbers (right plots) for {\sc Phantom}. The top row shows the averaged numbers for particles which have $\nabla \cdot \bm{v} < 0$, while the bottom row is averaged for regions where $\rho > 10 \rho_0$. The higher density regions have approximately double the kinetic and magnetic Reynolds numbers. The drop in Reynolds and Prandtl numbers over time is due to the fast MHD wave speed increasing in the signal velocity of the artificial dissipation terms. The Prandtl numbers are about unity, though decrease with resolution.}
\label{fig:ReRmPrandtl}
\end{figure*}

\subsection{Prandtl numbers in Eulerian schemes}

The primary source of numerical dissipation in Eulerian schemes is from the discretisation of advection terms. Consider a simple example of the contents of one grid cell advecting into an adjacent grid cell. If only a partial amount is transferred into the adjacent cell, then the contents must be reconstructed from the flux across the boundary. This approximation introduces diffusion due to its truncation error \citep[e.g.,][]{robertsonetal10}. The diffusion term in the first-order upwind scheme of \citet{cir52}, for example, scales according to $\propto v \Delta x (1 - \vert C \vert)$, where $C=v \Delta t / \Delta x$ is the Courant number. Higher order methods will change the scaling of the diffusion, but in all schemes it depends upon the resolution, time step size, and fluid velocity. 

Quantifying the effective numerical dissipation may be done by comparing simulations against analytic solutions. \citet{lb07} compared the analytic solution of a linear mode of the magneto-rotational instability (MRI) to shearing box simulations in order to calibrate their version of {\sc Zeus3D}. They varied the size of the time step and investigated resolutions from $32^3$ to $128^3$, determining that the total numerical dissipation (viscous and resistive) scaled linearly with time and quadratically with resolution. They found the magnetic Prandtl number to be approximately 2 (though in the context of this comparison, these simulations are for subsonic flows). In a similar manner, \citet{fromangetal07} performed simulations of the MRI with and without physical viscous and resistive dissipation terms. They found that the results of their ideal MHD simulations (dissipation is purely numerical) corresponded to ${\rm Pm}\approx2$, though cautioned that this depends upon the nature of the flow.

The effective Prandtl number for the version of {\sc Flash} used in this paper was calibrated by \citet{federrathetal11}. Using simulations of the small-scale dynamo amplification of a magnetic field, they compared results from ideal MHD simulations to simulations employing a fixed dissipation (at varying resolution). They found that ${\rm Pm}\approx2$ for flows of Mach numbers $0.4$ and $2$. Thus, it is expected that the {\sc Flash} calculations in our comparison will have a similar Prandtl number.

\subsection{Prandtl numbers in SPMHD}

In SPMHD, the equations of motion are derived from the discretised Lagrangian \citep{pm04b, price12}. Advection is computed exactly. Hence, the only sources of numerical dissipation are from the explicit sources of artificial viscosity and resistivity, which can be used to estimate the Reynolds and Prandtl numbers.

Artificial viscosity and resistivity in SPMHD are discretisations of physical dissipation terms, but with diffusion parameters that depend on resolution. \citet{al94} and \citet{murray96} analytically derived the amount of corresponding physical dissipation from the \citet{mg83} form of artificial viscosity (see also \citealt{monaghan05, lp10}). The artificial viscosity acts as both a shear and bulk viscosity. In these calculations, we use the \citet{monaghan97} form of artificial viscosity, which is similar except for the absence of a factor $h / \vert r_{ab} \vert$. \citet{mb12} calculated the amount of viscosity this adds in the continuum limit, and have shown that for the \citet{monaghan97} form of viscosity, it is approximately 18\% stronger for the $\alpha$ term. Using this approach, they also derived the coefficients for the $\beta_{\rm AV}$ term in the signal velocity. The shear viscosity in our simulations is estimated according to
\begin{equation}
\nu_{\rm AV} = \frac{62}{525} \alpha v_{\rm sig} h + \frac{9}{35 \pi} \beta_{\rm AV} \vert \nabla \cdot \bm{v} \vert h^2.
\label{eq:physicalnu}
\end{equation}
where
\begin{equation}
v_{\rm sig} = \sqrt{c_{\rm s}^{2} + v_{\rm A}^{2}}.
\end{equation}
The bulk viscosity will be $5/3\times$ this value \citep{lp10}. 

These coefficients are twice the values quoted by \citet{mb12}. Their work is derived in the context of a Keplerian accretion disc, in which they safely assume that half the particles inside a particle volume are approaching while the other half are receding. It is standard in SPH to apply artificial viscosity only to approaching particles. In this paper, we calculate Reynolds numbers for particles where $\nabla\cdot\bm{v}<0$, and use the full value of the coefficient as it is expected that inside a shock, nearly all particles will be approaching. In order to compare Reynolds numbers with {\sc Flash}, we compute the Reynolds numbers using only the shear viscosity, but note that the artificial viscosity will introduce a bulk viscosity which would be dynamically relevant for supersonic flows. Including this term would reduce the estimate of the Reynolds number and lead to larger ${\rm Pm}$ values, but it is not clear how bulk viscosity would affect the small-scale dynamo.

The corresponding physical dissipation from the artificial resistivity can be calculated in a similar manner \citep{price12}. Artificial resistivity corresponds to a physical resistivity given by
\begin{equation}
\eta_{\rm AR} \approx \frac{1}{2} \alpha_{\rm B} v_{\rm sig} h.
\label{eq:eta_AR}
\end{equation}
We note, as concluded in \citet{tp13}, that the $\beta_{\rm AV}$ term in the artificial viscosity is not required for artificial resistivity. It is added to artificial viscosity to prevent particle interpenetration in high Mach number shocks, and otherwise leads to unnecessary dissipation if added to artificial resistivity.

Since the dissipation terms use the local signal velocity, and our simulations use switches to dynamically adjust the values of $\alpha$ and $\alpha_{\rm B}$ for each particle, $\nu_{\rm AV}$ and $\eta_{\rm AR}$ are calculated per particle. In Fig.~\ref{fig:ReRmPrandtl}, we show the average kinetic Reynolds, magnetic Reynolds, and magnetic Prandtl numbers on the particles for our simulations. We find that the mean Prandtl number in these set of SPMHD calculations is approximately unity. The Prandtl number decreases with resolution, a consequence of the quadratic scaling of the $\beta_{\rm AV}$ term, which is present in the artificial viscosity but not artificial resistivity. The Prandtl number also decreases with time. This results from the signal velocity scaling behaviour, as the dissipation from the $\alpha$ term increases as the magnetic field is amplified ($v_{\rm A}$ increasing). The $\beta_{\rm AV}$ term is unaffected by this. For high-density regions ($\rho > 10$), we note that the Reynolds numbers are increased by approximately a factor of 2, directly corresponding to the reduction in $h$.

\bsp	
\label{lastpage}
\end{document}